\documentclass[nofootinbib,10pt,prb,aps,showpacs,twocolumn,article]{revtex4}
\usepackage[caption=false]{subfig}
\usepackage{amssymb}
\usepackage{amsmath}
\usepackage{commath}
\usepackage{mathtools}
\usepackage{graphicx,bm}
\usepackage{verbatim}
\usepackage{color}
\usepackage{pdfpages}
\usepackage{hyperref}
\usepackage{bbold}
\usepackage[normalem]{ulem} % to cross out text
\usepackage{blindtext}
\usepackage{verbatim}
\usepackage{relsize}
\def\be#1\ee{\begin{equation}#1\end{equation}}
\def\ba#1\ea{\begin{align}#1\end{align}}

\newcommand{\eqqref}[1]{Eq.\,\eqref{#1}}	
\newcommand{\figgref}[1]{Fig.\,\ref{#1}}

\newcommand{\bk}	{{\bm k}}

\newcommand{\bq}	{{\bm q}}

\newcommand{\bx}	{{\bm x}}
\newcommand{\bt}    {{\bm t}}

\newcommand{\bE}    {{\bm E}}
\newcommand{\bA}	{{\bm A}}
\newcommand{\bw}	{{\bm \omega}}

\newcommand{\bnabla}{{\bm \nabla}} 
\newcommand{\bJ}	{{\bm J}}

\newcommand{\scme}  {{\s^{\text{CME}}}}

\newcommand{\sahe} {{\s^{\text{AHE}}}}
\newcommand{\ane}  {\alpha^{\text{ANE}}}
\newcommand{\kthe}  {\kappa^{\text{THE}}}
\newcommand{\wmin}  {\omega_{\text{min}}}
\newcommand{\wmax}  {\omega_{\text{max}}}

\newcommand{\intdfour}[1]  {\int\!\!\frac{\text{d}^4 #1}{(2\pi)^4}}

\newcommand{\g}[1]	{\gamma^{#1}}
\newcommand{\G}[1]	{\Gamma^{#1}}

\newcommand{\eps}	{\epsilon}

\newcommand{\s}	    {\sigma}
\newcommand{\w}	    {\omega}

\newcommand{\veps}  {\varepsilon}

\newcommand{\NFD}[1]			   {N_{\text{F}}(#1)}

\newcommand{\mux}{\mu_{\chi}}

\newcommand{\tpara}{t_{\parallel}}
\newcommand{\bpsi}{\bar{\psi}}

\begin{document}
\title{Magnetovortical and thermoelectric transport in tilted Weyl metals}
\author{E.C.I. van der Wurff}
\email{e.c.i.vanderwurff@uu.nl}
\author{H.T.C. Stoof}
\email{H.T.C.Stoof@uu.nl}
\affiliation{Institute for Theoretical Physics, Utrecht University, Princetonplein 5, 3584 CC Utrecht, The Netherlands}
\date{\today}
\begin{abstract}
We investigate how tilting affects the off-diagonal, dissipationless response of a pair of chirally imbalanced Weyl cones to various external perturbations. The pair of chirally imbalanced Weyl cones can be described as a chiral electron fluid, that can flow with a velocity field that contains vorticity. Upon applying an external magnetic field, we obtain the so-called magnetovortical linear-response matrix that relates electric and heat currents to the magnetic field (chiral magnetic effect) and the vorticity (chiral vortical effect). We show how this reponse matrix becomes anisotropic upon tilting the cones and determine its non-analytic long-wavelength behavior, as well as the corresponding AC response. In addition, we discuss how the tilt dependence of the electronic (or density-density) susceptibility introduces anisotropy in the dispersion relation of the sound-like excitations in the fluid of chiral fermions, which are known as chiral magnetic waves. In the case of an externally applied electric field and a temperature gradient, we find a Hall-like response in the electric and heat current density that is perpendicular to both the tilting direction and the perturbations. As the tilting direction forms a time-reversal symmetry breaking vector, a non-zero (heat) orbital magnetization manifests itself. We calculate the magnetization currents microscopically and elucidate how to subtract these contributions to obtain the transport currents. 
\end{abstract}

%71.55.Ak 	Metals, semimetals, and alloys
%72.15.Jf 	Thermoelectric and thermomagnetic effects
%72.15.Gd Galvanomagnetic and other magnetotransport effects 
\maketitle
\section{Introduction}
The most important symmetry principle of particle physics is Lorentz invariance. Indeed, requiring invariance under Lorentz transformations yields a powerful restriction on which equations are eligible to describe the particles we encounter in Nature. For instance, it was Lorentz invariance, together with the wish for a counterpart to the Schr\"odinger equation that was first order in time derivatives, that allowed Paul Dirac to derive his famous equation describing massive spin-$1/2$ particles in 1928 \cite{Dirac1928}. The price that Dirac had to pay for finding an equation that obeyed these two requirements was that the spin-$1/2$ particle had to be described in terms of a four-component spinor, instead of the expected two-component wavefunction. It was only one year later when Hermann Weyl realized that Dirac's equation simplified greatly when considering massless spin-$1/2$ particles \cite{Weyl1929}. Instead of one equation involving a four-component spinor, Weyl obtained two decoupled equations, each for a two-component spinor with a definite chirality. Weyl fermions were, at least theoretically, born.

Contrastingly, not even translational symmetry is fully preserved in the presence of the atomic lattice out of which any ordinary solid is built up. Invariance under the even bigger Lorentz group thus seems too much to ask for in condensed matter. However, in certain cases Lorentz invariance can emerge at low energies in solid-state materials. One example of such a case occurs in the recently discovered Weyl semimetals \cite{Savrasov2011,Hasan2015A,Hasan2015B,Hasan2015C,Soljacic2015,Lv2015A,Lv2015B}. These materials host quasiparticles in their low-energy bandstructure that obey the aforementioned Weyl equation. This leads to a conical dispersion relation just like the light cone for massless particles known from particle physics, albeit with the speed of light replaced by the, typically much smaller, Fermi velocity. 

These so-called Weyl cones are topological: depending on the chirality of the cone, they acts as a sink or drain of Berry curvature in momentum space \cite{Volovik2009}. Only when the distance in energy-momentum space between two Weyl nodes with opposite chirality becomes zero, the monopoles annihilate, yielding a doubly-degenerate Dirac cone. Reversely, a pair\cite{Ninomiya1981} of non-degenerate Weyl cones can emerge from a doubly-degenerate Dirac cone in two distinct ways. Breaking time-reversal symmetry yields two Weyl cones separated in momentum space, whereas breaking of inversion symmetry yields two Weyl cones separated in energy space\cite{Burkov2011,Burkov2012}.

Interestingly, the emergent Lorentz symmetry in Weyl semimetals is not enforced by any crystal symmetries and thus generically it will be broken. The simplest way for this to happen is when the cones are tilted \cite{Goerbig2008, Bergholtz2015}. This is achieved mathematically by adding a term to the low-energy Hamiltonian that is proportional to the unit matrix in spin space and linear in momentum. Cones that are only slightly tilted are referred to as type-I Weyl cones, whereas cones that are tipped over are called type-II Weyl cones\cite{Bernevig2015,Bernevig2016A}. The existence of such tilted Weyl cones raises many interesting questions. For instance, we can ask how the diagonal optical response to an electric field is altered by the tilt\cite{Carbotte2016,Fritz2017B}, how the renormalization-group flow equations change \cite{Vozmediano2018}, what happens to the Landau level structure\cite{Goerbig2017}, how do tilted Weyl cones respond to disorder \cite{Bergholtz2017,Fritz2017A} and finally we can even show that vertex corrections due to Coulomb interactions naturally tilt the cone in the presence of a magnetic field\cite{Wurff2016}. 

In this paper we discuss how tilting the cones affects the electric and thermal transport of a Weyl metal. More specifically, we focus on the off-diagonal, dissipationless transport. It is important to note that the tilting direction of the Weyl cones forms another time-reversal symmetry breaking vector, besides the displacement vector in momentum space connecting the two Weyl cones. The thermoelectric response driven by an external electric field ${\bf E}$ and a thermal gradient ${\bm \nabla} T$ therefore contains a Hall part, describing currents that are perpendicular to the tilting direction\cite{Bardarson2017B,Wurff2017,Pesin2017}. Besides this novel thermoelectric response, we show that the magnetovortical response to an externally applied magnetic field ${\bf B}$ and a vorticity $\bw$ becomes anisotropic due to the tilting of the cones. Furthermore, we discuss the nonanalytic frequency-momentum behavior of the various transport coefficients in detail. 

This paper is organized as follows. We introduce the minimal model for a chirally imbalanced Weyl metal with tilted cones in Sec.\,\ref{sec:offdiagonal}. Subsequently we discuss the magnetovortical and thermoelectric response of a material that is described by such a Hamiltonian and give a summary of the main results we have obtained. The rest of the paper is devoted to a more in-depth discussion of the various properties of the transport coefficients. We explain how to use linear-response theory to calculate the transport coefficients due to the perturbations ${\bf B}$, ${\bf E}$, ${\bm \nabla} T$ and ${\bm \omega}$ in Sec.\,\ref{sec:linres}. Next, we calculate and discuss the tilt-dependent, anisotropic magnetovortical effects in Sec.\,\ref{sec:chiral} in three different regimes: the long-wavelength limit (Sec.\,\ref{subsec:lwl}), the static and homogeneous limit (Sec.\ref{subsec:stathomg}) and finally the AC frequency response (Sec.\,\ref{subsec:AC}). In Sec.\,\ref{sec:transport} we obtain the thermoelectric response due to tilted cones by explicitly calculating the magnetization contributions to the currents and subtracting them. Finally, we discuss our results in Sec.\,\ref{sec:discussion}.
%%%%%%%%%%%%%%%%%%%%%%%%%%%%%%%%%%%%%%%%%%%%%%%%%%%%%%%%%%%%%%%%%%%%%%%%%%%%%%%%%%%%%%%%%%%%%%%%%%%%%%%%%%%%%%%%%%%%%%%%%%%%%%%%%%%%%%%%%%%%%%%%%%%%%%%%%%%%%%%%%%%%%%%%%%%%%%%%%%%%%%%%%%%%%%%%%%%%%%%%%%%%%%%%%%%%%%%%%%%%%%%%%%%%%%%%%%%%%%%%%%%%%%%%%%%%%%%%%%%%%%%%%%%%%%%%%%%%%%%%%%%%%%%%%%%%%%%%%%%%%%%%%%%%%%%%%%%%%%%%%%%%%%%%%%%%%%%%%%%%%%%%%%%%%%%%%%%%%%%%%%%%%%%%%%%%%%%%%%%%%%%%%%%%%%%%%%%%%%%%%%%%%%%%%%%%%%%%
\section{Dissipationless transport} \label{sec:offdiagonal}
In this section we start by introducing a minimal model for a chirally imbalanced Weyl metal with tilted cones. Subsequently, we discuss the magnetovortical and thermoelectric response and highlight what changes upon tilting the cones, thereby summarizing the main results of the rest of the paper.
%%%%%%%%%%%%%%%%%%%%%%%%%%%%%%%%%%%%%%%%%%%%%%%%%%%%%%%%%%%%%%%%%%%%%%%%%%%%%%%%%%%%%%%%%%%%%%%%%%%%%%%%%%%%%%%%%%%%%%%%%%%%%%%%%%%%%%%%%%%%%%%%%%%%%%%%%%%%%%%%%%%%%%%%%%%%%%%%%%%%%%%%%%%%%%%%%%%%%%%%%%%%%%%%%%%%%%%%%%%%%%%%%%%%%%%%%%%%%%%%%%%%%%%%%%%%%%%%%%%%%%%%%%%%%%%%%%%%%%%%%%%%%%%%%%%%%%%%%%%%%%%%%%%%%%%%%%%%%%%%%%%%%%%%%%%%%%%%%%%%%%%%%%%%%%%%%%%%%%%%%%%%%%%%%%%%%%%%%%%%%%%%%%%%%%%%%%%%%%%%%%%%%%%%%%%%%%%%
\subsection{Model for tilted Weyl cones} \label{sec:model}
\begin{figure}[t]
\includegraphics[scale=.80]{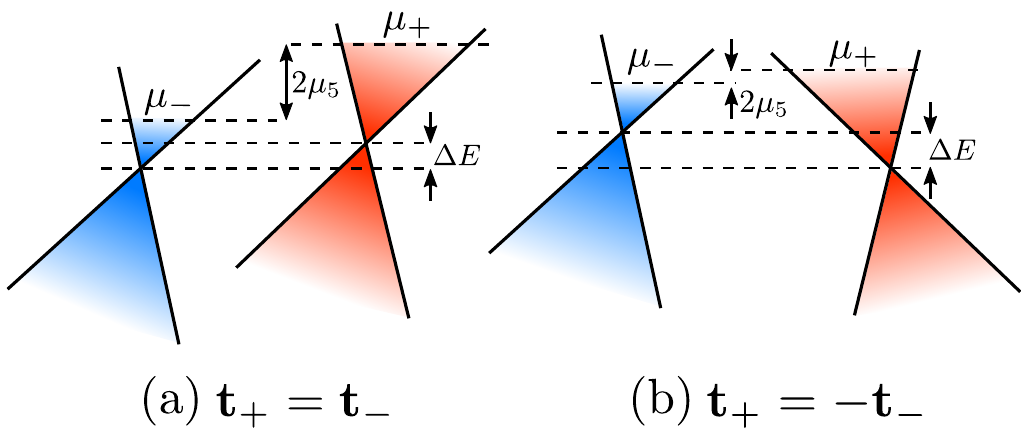}
\caption{Schematic depiction of a pair of tilted Weyl cones with negative (positive) chirality in blue (red) and corresponding chemical potential $\mu_-$ ($\mu_+$). The chiral imbalance is $\mu_5 = (\mu_+ - \mu_-)/2$ and the nodes are separated by an energy difference $\Delta E$. In this paper we mostly discuss the case $\Delta E = 0$. (a) Inversion-symmetry breaking tilt: Weyl cones are tilted by the same amount in one direction. (b) Inversion-symmetry retaining tilt: Weyl cones are tilted by the same amount in opposite directions. Note that we defined the tilting direction such that the associated energy contribution in \eqqref{eq:Ham} increases for momenta in the direction of the tilt.} \label{fig:Cones}
\end{figure}

We consider a doped time-reversal symmetry breaking Weyl metal. Because we focus on the off-diagonal response due to the tilting of the cones, we do not take an explicit separation between the Weyl nodes into account. Then, the simplest continuum two-band, grand-canonical Hamiltonian describing a tilted Weyl cone with chirality $\chi = \pm$ and isotropic\footnote{In principle, there can be different Fermi velocities in all three directions. This anisotropy can however always be transformed away by an appropriate scaling of the momenta.} Fermi velocity $v_F$ is given by
\be
\mathcal{H}_{\chi}(\bk) = \chi \hbar v_F \bk \cdot {\bm \sigma} + \big(\hbar v_F  \bk \cdot \bt_{\chi} - \mux\big)\sigma^0, \label{eq:Ham}
\ee
with $\s^{\mu} = (\mathbb{1}_2, \bm \s)$ the four-vector of Pauli matrices and $\mu_{\pm} \equiv \mu \pm \mu_5$ the chemical potential of the Weyl node with chirality $\pm$, in terms of the chemical potential $\mu \equiv (\mu_+ + \mu_-)/2$ and the chiral, or axial, chemical potential $\mu_5\equiv (\mu_+ - \mu_-)/2 $. Furthermore, the tilting direction of each cone is indicated by $\bt_{\chi}$. For simplicity we take $0 \leq |\bt_{\chi}| = t < 1$, meaning that we consider type-I Weyl cones that are tilted by the same amount. The latter requirement is easily generalized if necessary. Next to the magnitude, each cone can also have a different tilting direction. Indeed,  ${\bf t}_{\chi} = \chi {\bf t}$ is the inversion-symmetric case and ${\bf t}_{\chi} = {\bf t}$ when inversion symmetry is broken. Physically, the inversion-symmetric case corresponds to the situation where the two Weyl cones are tilted in opposite directions by exactly the same amount. This is pictorially displayed in \figgref{fig:Cones}. 
%%%%%%%%%%%%%%%%%%%%%%%%%%%%%%%%%%%%%%%%%%%%%%%%%%%%%%%%%%%%%%%%%%%%%%%%%%%%%%%%%%%%%%%%%%%%%%%%%%%%%%%%%%%%%%%%%%%%%%%%%%%%%%%%%%%%%%%%
%%%%%%%%%%%%%%%%%%%%%%%%%%%%%%%%%%%%%%%%%%%%%%%%%%%%%%%%%%%%%%%%%%%%%%%%%%%%%%%%%%%%%%%%%%%%%%%%%%%%%%%%%%%%%%%%%%%%%%%%%%%%%%%%%%%%%%%%
For later reference, we note that the eigenvalues of the Hamiltonian in \eqqref{eq:Ham}, which are only valid for $\Delta E = 0$ (c.f. \figgref{fig:Cones}), are given by
\be
\veps_{n\bk} -\mux = n\hbar v_F|\bk| + \hbar v_F\bk\cdot\bt_{\chi} - \mux, \label{eq:energy}
\ee
with $n = \pm$ indicating the conduction (+) and valence (-) band, respectively. 

The presence of a chiral imbalance, indicated in \eqqref{eq:Ham} by $\mu_+ \neq \mu_-$, is a non-equilibrium property. One way to generate a chiral imbalance is by irrediating a Weyl semimetal with circularly polarized light, thereby transferring chirality from the light to the electrons \cite{Kharzeev2019}. Another way is to apply strain to the Weyl semimetal \cite{Kharzeev2016}. We instead focus on the possibility to pump charge from one cone to the other by applying non-orthogonal electric and magnetic fields, which we discuss in the next section. Whatever the pumping mechanism is, it will be counterbalanced by an intervalley scattering time $\tau_5$ that inevitably is present. In the end, a steady state develops, which we took as a starting point in \eqqref{eq:Ham}. 
%%%%%%%%%%%%%%%%%%%%%%%%%%%%%%%%%%%%%%%%%%%%%%%%%%%%%%%%%%%%%%%%%%%%%%%%%%%%%%%%%%%%%%%%%%%%%%%%%%%%%%%%%%%%%%%%%%%%%%%%%%%%%%%%%%%%%%%%%%%%%%%%%%%%%%%%%%%%%%%%%%%%%%%%%%%%%%%%%%%%%%%%%%%%%%%%%%%%%%%%%%%%%%%%%%%%%%%%%%%%%%%%%%%%%%%%%%%%%%%%%%%%%%%%%%%%%%%%%%%%%%%%%%%%%%%%%%%%%%%%%%%%%%%%%%%%%%%%%%%%%%%%%%%%%%%%%%%%%%%%%%%%%%%%%%%%%%%%%%%%%%%%%%%%%%%%%%%%%%%%%%%%%%%%%%%%%%%%%%%%%%%%%%%%%%%%%%%%%%%%%%%%%%%%%%%%%%%%
\subsection{Magnetovortical response}
An interesting property of the quantum theory of Weyl fermions is that the amount of left-handed and right-handed Weyl fermions is not separately conserved when subjected to externally applied magnetic and electric fields that are non-orthogonal. In vacuum this leads to an interesting paradox: particles of one chirality seem to disappear, whereas particles with the opposite chirality appear out of nothing. This phenomenon is called the chiral anomaly and it is proportional to ${\bf E} \cdot {\bf B}$\cite{Jackiw1969,Adler1969}. In a condensed-matter system hosting Weyl cones the explanation of the chiral anomaly is straightforward. Namely, in a bandstructure the Weyl cones always come in pairs that are connected via the rest of the bandstructure \cite{Ninomiya1981}. Applying external electric and magnetic fields will subsequently transfer population from one cone to the other, thereby converting the quasiparticles from one type of chirality into the other. The result is a chiral imbalance, which is signaled by a distinct chemical potential $\mu_{\pm}$ for the cone with chirality $\pm$, which we already used in \eqqref{eq:Ham}.

It is exactly this chiral imbalance that gives rise to interesting transport properties in the presence of an external magnetic field ${\bf B}$ and a vorticity $\bw = ({\bm \nabla} \times {\bm v})/2$ due to a non-zero local velocity ${\bm v}$ of the fermion fluid\cite{Landsteiner2011}. The most famous of these magnetovortical effects is the chiral magnetic effect (CME) \cite{Ninomiya1983,Burkov2012,Kharzeev2006,Warringa2009}, which constitutes an electric current density in the direction of an externally applied magnetic field: $\langle {\bm J}_e \rangle = \s^{\text{CME}}{\bf B}$. The coupled magnetovortical response for the electric and energy current densities $\langle \bJ_e \rangle$ and $\langle \bJ_{\veps}\rangle$ is neatly summarized in the response matrix
\be
\renewcommand\arraystretch{1.3}
\begin{pmatrix} \langle {\bm J}_{e} \rangle \\ \langle {\bm J}_{\veps} \rangle \end{pmatrix} 
= 
\begin{pmatrix} \s^{\text{CME}} & \s^{\text{CVE}} \\ \s_{\veps}^{\text{CME}} & \s_{\veps}^{\text{CVE}} \end{pmatrix} 
\begin{pmatrix} {\bf B} \\ 2\bw/v_F^2 \end{pmatrix}, \label{eq:Transport1a}
\ee
with $\s^{\text{CVE}}$, $\s^{\text{CME}}_{\veps}$ and $\s^{\text{CVE}}_{\veps}$ the linear-response transport coefficients of the magnetovortical effects (CME/CVE) \cite{Vilenkin1978,Vilenkin1979}. This response matrix clarifies why the magnetovortical response is dissipationless. The pseudovectors ${\bf B}$ and $\bm{\omega}$ are odd under time reversal, just like the electric and energy current densities. The real part of the corresponding conductivities is therefore necessarily even under time reversal, signaling that the conductivities cannot be due to dissipation and hence warrant the name dissipationless.

The response matrix in \eqqref{eq:Transport1a} is written in terms of the electric and energy current density. In condensed matter, however, the natural reference energy for the carriers of charge and energy is the chemical potential. It is therefore customary to define the heat current density 
\be
\langle \bJ_Q \rangle \equiv \sum_{\chi = \pm} \Big[\langle \bJ^{\chi}_{\veps}\rangle + \frac{\mux}{e}\langle \bJ^{\chi}_e \rangle\Big],
\ee
with $-e$ the electron charge. The associated response matrix reads
\be
\renewcommand\arraystretch{1.3}
\begin{pmatrix} \langle \bJ_{e} \rangle \\ \langle \bJ_{Q} \rangle \end{pmatrix} 
=
\begin{pmatrix} \s^{\text{CME}} & \s^{\text{CVE}} \\ \s_{Q}^{\text{CME}} & \s_{Q}^{\text{CVE}} \end{pmatrix} 
\begin{pmatrix} {\bf B} \\ 2\bw/v_F^2 \end{pmatrix}, \label{eq:Transport1b}
\ee
with 
\begin{subequations}
\ba
\s_{Q}^{\text{CME}} &\equiv \sum_{\chi=\pm}\Big[\s_{\veps,\chi}^{\text{CME}} + \frac{\mux}{e}\s_{\chi}^{\text{CME}}\Big], \\
\s_{Q}^{\text{CVE}} &\equiv \sum_{\chi=\pm}\Big[\s_{\veps,\chi}^{\text{CVE}} + \frac{\mux}{e}\s_{\chi}^{\text{CVE}}\Big].
\ea
\end{subequations}
All magnetovortical effects are only non-zero in the presence of a non-zero chiral imbalance and the off-diagonal elements of \eqqref{eq:Transport1b} also require a non-zero chemical potential, as we show later. There is a similar response matrix for the axial currents \cite{Landsteiner2013}, which are defined as the difference between, instead of the sum of, the currents coming from the separate Weyl cones. In this case the corresponding transport coefficients all require a non-zero chemical potential $\mu$ and the off-diagonal components in \eqqref{eq:Transport1b} again require a non-zero chiral imbalance $\mu_5$ as well. We, however, do not pursue this direction here, although our methods can easily be used in this case as well because the response is diagonal in the chirality\cite{Landsteiner2014A}. 

We instead investigate how the magnetovortical transport changes upon tilting the cones. Due to the fact that the tilting direction breaks rotational symmetry, the response matrix from \eqqref{eq:Transport1b} is generalized to 
\be
\renewcommand\arraystretch{1.3}
\begin{pmatrix} \langle J^i_{e} \rangle \\ \langle J^i_{Q} \rangle \end{pmatrix} 
=
\begin{pmatrix} \s_{ij}^{\text{CME}} & \s_{ij}^{\text{CVE}} \\ \s_{Q,ij}^{\text{CME}} & \s_{Q,ij}^{\text{CVE}} \end{pmatrix} 
\begin{pmatrix} B^j \\ 2\w^j/v_F^2 \end{pmatrix}. \label{eq:Transport1c}
\ee
As it turns out the chiral magnetic conductivities remain isotropic, i.e., $\sigma_{ij}^{\text{CME}} = \sigma^{\text{CME}}\delta^{ij}$ and $\sigma_{Q,ij}^{\text{CME}} = \sigma^{\text{CME}}_Q\delta^{ij}$, which is a property that is ultimately enforced by the chiral anomaly. The vortical conductivities, on the other hand, become anisotropic and can be decomposed in a transverse and longitudinal part with respect to the tilting direction $\hat{t}^i = t^i/t$ as follows
\be
\s_{ij}^{\text{CVE}} = \sigma^{\text{CVE}}_{\perp}\big(\delta_{ij} - \hat{t}_i\hat{t}_j\big) + \sigma^{\text{CVE}}_{\parallel}\hat{t}_i\hat{t}_j, \label{eq:CVElongtrans}
\ee
and analogous expressions hold for the chiral vortical heat or energy conductivities.

We calculate all the magnetovortical transport coefficients in Sec.\ref{sec:chiral} in the long-wavelength limit and in addition obtain their frequency dependence. For the static and homogeneous limit we summarize all results obtained in Table \ref{tab:magnetovortical}. In Sec.\ref{subsec:stathomg} we comprehensively discuss all results presented there and give physical arguments to clarify them.
%%%%%%%%%%%%%%%%%%%%%%%%%%%%%%%%%%%%%%%%%%%%%%%%%%%%%%%%%%%%%%%%%%%%%%%%%%%%%%%%%%%%%%%%%%%%%%%%%%%%%%%%%%%%%%%%%%%%%%%%%%%%%%%%%%%%%%%%%%%%%%%%%%%%%%%%%%%%%%%%%%%%%%%%%%%%%%%%%%%%%%%%%%%%%%%%%%%%%%%%%%%%%%%%%%%%%%%%%%%%%%%%%%%%%%%%%%%%%%%%%%%%%%%%%%%%%%%%%%%%%%%%%%%%%%%%%%%%%%%%%%%%%%%%%%%%%%%%%%%%%%%%%%%%%%%%%%%%%%%%%%%%%%%%%%%%%%%%%%%%
\subsection{Thermoelectric transport}
As alluded to before, Weyl cones also responds interestingly in the presence of an electric field $\bf E$ and a temperature gradient ${\bm \nabla} T$, even in the absence of a chiral imbalance. In the case of a time-reversal symmetry breaking Weyl semimetal with two Weyl cones separated in momentum space, there is a topological off-diagonal response. Most famously, for non-tilted Weyl cones the associated intrinsic topological anomalous Hall effect (AHE) is given by \cite{Burkov2011}   
\be
\langle {\bm J}_e \rangle = \frac{e^2}{4\pi^2\hbar}\Delta \bk\times {\bf E}, \label{eq:AHEintro}
\ee
with $\Delta\bk$ the momentum-space separation between the Weyl nodes. Furthermore, the breaking of time-reversal symmetry allows for a topological thermal Hall effect (THE), which is the flow of a transverse heat current as a response to a temperature gradient in the absence of an electric current\cite{Tewari2013}, i.e.,
\be
\langle {\bm J}_Q \rangle  = -\frac{k_B^2T}{12\hbar} \Delta\bk\times{\bm \nabla}T. \label{eq:THEintro}
\ee
Due to the existence of thermoelectricity, we may also expect a transverse electric current due to a temperature gradient. Such an anomalous Nernst effect (ANE) is however absent in a simple, linear continuum model of the Weyl cones\cite{Fiete2014}. Furthermore, we note that besides the intrinsic, topological contributions presented in Eqs.\,\eqref{eq:AHEintro} and \eqref{eq:THEintro}, extrinsic contributions due to skew and side-jump scattering of electrons off impurities may also be present\cite{Niu2010}. From now on, however, we only consider the intrinsic, non-topological contributions to the thermoelectric response coefficients.

What happens to the thermoelectric response upon tilting the cones? Firstly, the transport coefficients are renormalized by the tilt \cite{Carbotte2016,Sukhachov2017}. Secondly, the tilt introduces another time-reversal symmetry-breaking vector,  
thereby allowing for a different contribution to the anomalous Hall effect\cite{Tiwari2016,Pesin2017,Wurff2017,Carbotte2018} in \eqqref{eq:AHEintro}, and to the thermal Hall effect\cite{Bardarson2017B} in \eqqref{eq:THEintro}. Additionally, the anomalous Nernst effect becomes non-zero, even in a linear model \cite{Bardarson2017B,Tewari2018}. The off-diagonal, explicitly tilt-dependent part of the response matrix for the total electric and heat current densities can then be written as
\be
\renewcommand\arraystretch{1.3}
\begin{pmatrix} \langle {\bm J}_{e} \rangle \\ \langle {\bm J}_{Q} \rangle \end{pmatrix} = 
\begin{pmatrix} \s^{\text{AHE}} & \alpha^{\text{ANE}} T \\ \alpha^{\text{ANE}} T & \bar{\kappa}^{\text{THE}}T \end{pmatrix}
\begin{pmatrix} \bt\times{\bf E} \\ \bt \times {\bm \nabla} T/T \end{pmatrix} \label{eq:Transport2},
\ee
with $\bt$ the tilting direction. Onsager reciprocity forces the off-diagonal elements of the response matrix to be the same, so $\s^{\text{ANE}}_{Q}\equiv \ane T$. In \eqqref{eq:Transport2}, we have defined the anomalous Nernst conductivity as by Niu \emph{et al.}\cite{Niu2006}. The anomalous Nernst coefficient can, however, also be defined as the steady-state constant of proportionality between the voltage difference due to the redistribution of charge caused by an applied temperature gradient in a sample without leads. The anomalous Nernst coefficient is then given by $\vartheta^{\text{ANE}} \equiv -\alpha^{\text{ANE}}/\s^{\text{AHE}}$. 

Thermal conductivity is defined as a heat current in the absence of a charge current, i.e. $\langle \bJ_{Q}\rangle = \kappa^{\text{THE}}{\bt \times \bm \nabla} T\big|_{\langle \bJ_e \rangle = {\bf 0}}$. In this manner, the thermal Hall coefficient is found from \eqqref{eq:Transport2} to be
\be
\kappa^{\text{THE}} = \bar{\kappa}^{\text{THE}} - \frac{T(\alpha^{\text{ANE}})^2}{\sahe}, \label{eq:kappaintro}
\ee
where we only took off-diagonal contributions into account. In principle the linear-response coefficients $\s^{\text{AHE}}$, $\alpha^{\text{ANE}}$ and $\bar{\kappa}^{\text{THE}}$ can be calculated by extracting from the off-diagonal part of the appropriate current-current correlators the contribution that is proportional to $\bt$. However, when calculating the thermal transport coefficients $\ane$ and $\bar{\kappa}^{\text{THE}}$ a problem arises: they contain terms that are dependent on the chemical potential $\mu$, but independent of the temperature $T$. From the response matrix in \eqqref{eq:Transport2} it is clear that such a term renders the zero-temperature limit ill-defined. The physical explanation is that tilting the cones generates a non-zero magnetization density $\bm M$ in the direction of the tilt $\bt$. Such a magnetization density in turn yields a circulating current of the form $\bm \nabla \times {\bm M}$, that gives a contribution to the transport coefficients coming from the Kubo formulas, but is unobservable with a transport measurement\cite{Halperin1997,Niu2006,Niu2011}. In order to calculate the transport current that can be measured in experiment, this superfluous term should therefore be subtracted. How this can be achieved is discussed in Sec.\,\ref{subsec:magncontrib}.

Having discussed the most important differences that occur in the magnetovortical and thermoelectric response due to tilting the Weyl cones, we now turn to a more in-depth discussion of how to calculate all the corresponding transport coefficients.
%%%%%%%%%%%%%%%%%%%%%%%%%%%%%%%%%%%%%%%%%%%%%%%%%%%%%%%%%%%%%%%%%%%%%%%%%%%%%%%%%%%%%%%%%%%%%%%%%%%%%%%%%%%%%%%%%%%%%%%%%%%%%%%%%%%%%%%%%%%%%%%%%%%%%%%%%%%%%%%%%%%%%%%%%%%%%%%%%%%%%%%%%%%%%%%%%%%%%%%%%%%%%%%%%%%%%%%%%%%%%%%%%%%%%%%%%%%%%%%%%%%%%%%%%%%%%%%%%%%%%%%%%%%%%%%%%%%%%%%%%%%%%%%%%%%%%%%%%%%%%%%%%%%%%%%%%%%%%%%%%%%%%%%%%%%%%%%%%%%%%%%%%%%%%%%%%%%%%%%%%%%%%%%%%%%%%%%%%%%%%%%%%%%%%%%%%%%%%%%%%%%%%%%%%%%%%%%%%%%%%%%%%%%%%%%%%%%%%%%%%%%%%%%%%%%%%%%%%%%%%%%%%%%%%%%%%%%%%%%%%%%%%%%%%%%%%%%%%%%%%%%%%%%%%%%%%%%%%%%%%%%%%%%%%%%%%%%%%%%%%%%%%%%%%%%%%%%%%%%%%%%%%%%%%%%%%%%%%%%%%%%%%%%%%%%%%%%%%%%%%%%%%%%%%%%%%%%%%%%%%%%%%%%%%%%%%%%%%%%%%%%%%%%%%
\section{Linear-response theory} \label{sec:linres}
In this section we set up the linear-response theory necessary to derive all the transport coefficients that we discussed in the previous sections. We start by writing the action corresponding to the Hamiltonian \eqqref{eq:Ham} and giving the corresponding Green's function or propagator. We proceed by deriving the electric, momentum and energy currents using the Hamiltonian in \eqqref{eq:Ham} and the corresponding action. Subsequently, we discuss how these three currents give rise to nine different current-current correlation functions, of which six are relevant for the response matrices in Eqs.\,\eqref{eq:Transport1a}, \eqref{eq:Transport1b} and \eqref{eq:Transport2} we set out to calculate. Finally, we derive and give explicit expressions for the antisymmetric part of the current-current correlation functions and discuss how they can be appropriately decomposed. In what follows we take $\hbar \equiv 1$ and $v_F\equiv 1$, only reinstating $v_F$ in our final results.
%%%%%%%%%%%%%%%%%%%%%%%%%%%%%%%%%%%%%%%%%%%%%%%%%%%%%%%%%%%%%%%%%%%%%%%%%%%%%%%%%%%%%%%%%%%%%%%%%%%%%%%%%%%%%%%%%%%%%%%%%%%%%%%%%%%%%%%%%%%%%%%%%%%%%%%%%%%%%%%%%%%%%%%%%%%%%%%%%%%%%%%%%%%%%%%%%%%%%%%%%%%%%%%%%%%%%%%%%%%%%%%%%%%%%%%%%%%%%%%%%%%%%%%%%%%%%%%%%%%%%%%%%%%%%%%%%%%%%%%%%%%%%%%%%%%%%%%%%%%%%%%%%%%%%%%%%%%%%%%%%%%%%%%%%%%%%%%%%%%%%%%%%%%%%%%%%%%%%%%%%%%%%%%%%%%%%%%%%%%%%%%%%%%%%%%%%%%%%%%%%%%%%%%%%%%%%%%%%%%%%%%%%%%%%%%%%%%%%%%%%%%%%%%%%%%%%%%%%%%%%%%%%%%%%%%%%%%%%%%%%%%%%%%%%%%%%%%%%%%%%%%%%%%%%%%%%%%%%%%%%%%%%%%%%%%%%%%%%%%%%%%%%%%%%%%%%%%%%%%%%%%%%%%%%%%%%%%%%%%%%%%%%%%%%%%%%%%%%%%%%%%%%%%%%%%%%%%%%%%%%%%%%%%%%%%%%%%%%%%%%%%%%%%%%
\subsection{Electronic action and Green's functions} \label{subsec:action}
For calculational purposes it is convenient to combine the two copies of the Hamiltonian from \eqqref{eq:Ham} for the two cones with opposite chirality into a Dirac-like action in terms of the four-component spinor $\psi$, i.e.,
\be
S_0[\psi,\bar{\psi}] = \int \mathrm{d}^4x\,\bar{\psi}\big[\!-i\G{\mu}\partial_{\mu} -\mu\g{0} - \mu_5 \g{0}\g{5}   \big]\psi, \label{eq:action}
\ee
where $\gamma^{\mu}$ are the usual gamma matrices\footnote{For the gamma matrices we use the representation $\g{0} = i\s^y \otimes \mathbb{1}_2$, $\g{j} = \s^x \otimes \s^j$ and $\g{5}=-\s^z \otimes \mathbb{1}_2$, such that they obey a Clifford algebra $\{\g{\mu},\g{\nu}\} = 2\eta^{\mu\nu}$, with $\eta_{\mu\nu} = \text{diag}(-,+,+,+)$ the mostly-plus Minkowski metric.}, $\bpsi = \psi^{\dagger}\g{0}$ and $\G{\mu} \equiv \g{\mu} + \g{0}t^{\mu}$ with $t^{\mu} = (0,{\bf t})$. The Feynman propagator defined by \eqqref{eq:action} is given in momentum space by
\ba
S_F(k) &\equiv i\big(k_{\mu}\G{\mu} - \mu\g{0} - \mu_5\g{0}\g{5}\big)^{-1} \nonumber\\
&= i \begin{pmatrix} G_-(k) & 0 \\ 0 & G_+(k)\end{pmatrix}\g{0}, \label{eq:Prop}
\ea
where we used the four-vector notation $k^{\mu} = (\w,\bk)$ and introduced the propagator $G_{\pm}(k)$ for a single cone with chirality $\pm$. The Matsubara Green's function associated with the latter is given by
\be
G_{\chi}(i\w_n,\bk) = \frac{1}{2} \sum_{u = \pm}\frac{\s^0 - \chi u\hat{\bk}\cdot{\bm{\s}}}{i\w_n + \mux  + \veps_{u\bk}}, \label{eq:Green}
\ee
in terms of the fermionic Matsubara frequencies $\w_n = (2n+1)\pi/\beta$ with $\beta \equiv 1/k_BT$ and $\hat{\bk} = \bk / |\bk|$.
%%%%%%%%%%%%%%%%%%%%%%%%%%%%%%%%%%%%%%%%%%%%%%%%%%%%%%%%%%%%%%%%%%%%%%%%%%%%%%%%%%%%%%%%%%%%%%%%%%%%%%%%%%%%%%%%%%%%%%%%%%%%%%%%%%%%%%%%%%%%%%%%%%%%%%%%%%%%%%%%%%%%%%%%%%%%%%%%%%%%%%%%%%%%%%%%%%%%%%%%%%%%%%%%%%%%%%%%%%%%%%%%%%%%%%%%%%%%%%%%%%%%%%%%%%%%%%%%%%%%%%%%%%%%%%%%%%%%%%%%%%%%%%%%%%%%%%%%%%%%%%%%%%%%%%%%%%%%%%%%%%%%%%%%%%%%%%%%%%%%%%%%%%%%%%%%%%%%%%%%%%%%%%%%%%%%%%%%%%%%%%%%%%%%%%%%%%%%%%%%%%%%%%%%%%%%%%%%%%%%%%%%%%%%%%%%%%%%%%%%%%%%%%%%%%%%%%%%%%%%%%%%%%%%%%%%%%%%%%%%%%%%%%%%%%%%%%%%%%%%%%%%%%%%%%%%%%%%%%%%%%%%%%%%%%%%%%%%%%%%%%%%%%%%%%%%%%%%%%%%%%%%%%%%%%%%%%%%%%%%%%%%%%%%%%%%%%%%%%%%%%%%%%%%%%%%%%%%%%%%%%%%%%%%%%%%%%%%%%%%%%%%%%%%%
\subsection{Electric current, energy current and momentum density} \label{subsec:currents}
In order to calculate the coupled response matrices of the Weyl cones, we need to couple the electrons i) to an external gauge field $A_{\mu}$; ii) to a temperature gradient ${\bm \nabla T}$ and iii) to the vorticity ${\bm \omega}$. The first of these three is achieved by applying the minimal-coupling procedure $\partial_{\mu}\psi \rightarrow \partial_{\mu}\psi + i e A_{\mu}\psi$ to the action in \eqqref{eq:action}. This yields a coupling of the form $J^{\mu}_e A_{\mu}$, with $J^{\mu}_e \equiv e\bpsi\G{\mu}\psi$ the electric current density in terms of the previously defined tilt-dependent vertex $\G{\mu}$. 

Secondly, we couple the fermions to a temperature gradient. This was pioneered by Luttinger\cite{Luttinger1964} using a fictitious gravitational potential. Assuming a 
homogeneous temperature $T$, perturbed by small spatial variations $\delta T(\bx)$, the temperature profile can be written as 
\be
T(\bx,t) = T + \delta T(\bx)e^{-i\w t}.
\ee
Such an inhomogeneous temperature can be shown to act as a perturbation on the Minkowski metric $\eta_{\mu\nu}$\cite{Herzog2009}. In linear response the modified metric $g_{\mu\nu}$ is found to be 
\ba
\mathrm{d}s^2 &= \eta_{\mu\nu}\mathrm{d}x^{\mu}\mathrm{d}x^{\nu}\! - 2 \frac{e^{-i\w t}}{i\w}\frac{\partial_j T}{T} \mathrm{d}x^j\mathrm{d}t \label{eq:metric},
\ea
such that the change in metric is $\delta g_{j0} = -e^{-i\w t}\partial_j T / i\w T$. Such metric fluctuations couple to the energy-momentum tensor $T^{\mu\nu}$ in the action as $\delta g_{\mu\nu}T^{\mu\nu}$, meaning that they act as a source for the energy-momentum tensor. The off-diagonal part of the metric in \eqqref{eq:metric} thus couples to the energy current density $J_{\veps}^{\mu} \equiv T^{\mu0}$, which is defined by the conservation law of energy, i.e.,
\be 
\partial_0 T^{00} + \partial_j T^{j0} \equiv \partial_t \mathcal{E} + {\bm \nabla} \cdot {\bm J}_{\veps}  =  0, \label{eq:consE}
\ee
with $\mathcal{E}$ the canonical energy density. Physically, the energy current density is simply given by the hermitian, symmetrized expression of energy times the velocity. Using the action from \eqqref{eq:action}, we find
\be
T^{\mu0} = \frac{i}{2}\Big[\partial_j \bpsi \G{j}\g{0}\G{\mu}\psi - \bpsi \G{\mu}\g{0}\G{j}\partial_j\psi\Big], \label{eq:Tmu0}
\ee
which depends on the tilt via $\G{\mu}$. Note that we can also write an expression for $T^{\mu0}$ in terms of only temporal derivatives by imposing the equations of motion. Our expression for $T^{\mu0}$ obeys the conservation law \eqqref{eq:consE} with the \emph{canonical} energy density $\mathcal{E}(x) \equiv T^{00}$. Similarly, we could derive the heat current density ${\bm J}_Q$ by using the grand-canonical Hamiltonian density, that follows from \eqqref{eq:action}, in the conservation law in \eqqref{eq:consE}.

Finally, we consider how to include a vorticity ${\bm \omega} = (\bnabla \times {\bm v})/2$. Giving the electron fluid a non-zero velocity ${\bm v}$ is achieved by performing a Galilean transformation on the Hamiltonian \eqqref{eq:Ham}: $\mathcal{H}_{\chi}(\bk) \rightarrow \mathcal{H}_{\chi}(\bk) - \bk\cdot{\bm v}$. Alternatively, to make the connection with the previous discussion, we can consider the velocity to be a perturbation on the metric, i.e., $\delta g_{0i} = v^i$. This part of the metric then couples to the momentum density $J_p^{i} \equiv T^{0 i}$, which obeys the conservation law
\be
\partial_0 T^{0i} + \partial_j \Pi^{ji} = 0, \label{eq:consP}
\ee
with $\Pi^{ji} = i\big[\bpsi \G{j} \partial^i \psi - \partial^i \bpsi \G{j}\psi\big]/2$ the stress tensor. The momentum density is given explicitly by
\be
J_p^{\mu} = T^{0\mu} = \frac{i}{2}\Big[\bpsi\g{0}\partial^{\mu}\psi - \partial^{\mu}\bpsi\g{0}\psi\Big], \label{eq:T0mu}
%= \sum_{\bk}\psi^{\dagger}_{\bk}k^{\mu}\psi_{\bk}. \label{eq:T0mu}
\ee
from which the total (center-of-mass) momentum follows by averaging the spatial part over the whole space, i.e.,
\be 
\int\text{d}\bx \, {\bJ}_p(\bx) = \sum_{\bk}\psi^{\dagger}_{\bk}\bk\psi_{\bk},
\ee
as expected. As can be seen from Eqs.\,\eqref{eq:Tmu0} and \eqref{eq:T0mu}, the energy-momentum tensor is not manifestly symmetric, which we discuss in more depth lateron. In principle it can be made symmetric by adding suitable boundary terms to the action. We, however, refrain from doing so because Eqs.\,\eqref{eq:Tmu0} and \eqref{eq:T0mu} are the physical, conserved currents that are determined by the equations of motion for the Dirac field and the conservation laws in \eqqref{eq:consE} and \eqqref{eq:consP}.

%%%%%%%%%%%%%%%%%%%%%%%%%%%%%%%%%%%%%%%%%%%%%%%%%%%%%%%%%%%%%%%%%%%%%%%%%%%%%%%%%%%%%%%%%%%%%%%%%%%%%%%%%%%%%%%%%%%%%%%%%%%%%%%%%%%%%%%%%%%%%%%%%%%%%%%%%%%%%%%%%%%%%%%%%%%%%%%%%%%%%%%%%%%%%%%%%%%%%%%%%%%%%%%%%%%%%%%%%%%%%%%%%%%%%%%%%%%%%%%%%%%%%%%%%%%%%%%%%%%%%%%%%%%%%%%%%%%%%%%%%%%%%%%%%%%%%%%%%%%%%%%%%%
\subsection{Current-current response functions and their decomposition} \label{sec:currentcurrent}
Having obtained the coupling between the fermions and the external perturbations, we can integrate out the fermions to arrive at the effective action for the external perturbations. This effective action is quadratic in the external gauge field $A_{\mu}$ and the fluctuation of the metric $\delta g_{\mu\nu}$. The electric (energy) current now follows in linear response from taking the functional derivative of the effective action with respect to the gauge field (metric fluctuation), leading to
\ba
\langle J_a^{\mu}(q) \rangle &= \Pi_{ae}^{\mu\nu}(q)A_{\nu}(q) + \Pi_{ap}^{\mu\nu}(q)\delta g_{0\nu}(q) \nonumber \\
&\phantom{=}+ \Pi^{\mu\nu}_{a\veps}(q) \delta g_{\nu0}(q), \label{eq:current}
\ea
with $a \in \{e,\veps\}$. To ensure causality the response functions $\Pi_{ab}^{\mu\nu}(q)$ are understood to be retarded. They can generically be written as
\ba
&\hspace{-.2cm}i\Pi^{\mu\nu}_{ab}(q) \equiv -\int\! \text{d}^4(x-y)\langle J^{\mu}_a(x) J^{\nu}_b(y) \rangle e^{-iq_{\mu}(x-y)^{\mu}} \nonumber\\
&\phantom{=}\,=\intdfour{k}\text{Tr}\big[ \Lambda^{\mu}_a(k,q) S_F(k+q)\Lambda^{\nu}_b(k,q) S_F(k) \big]  \label{eq:Pigeneric1},\!\!
\ea
where we omitted disconnected contributions and contributions from the (energy) magnetization, to which we return later. The electric, momentum and energy vertices are given by
\begin{subequations}
\ba
\Lambda^{\mu}_e &=e\G{\mu}, \label{eq:Vertexe}\\
\Lambda^{\mu}_p(k,q) &= -\frac{1}{2}\big(2k^{\mu} + q^{\mu}\big)\g{0}, \label{eq:Vertexp} \\
\Lambda^{\mu}_{\veps}(k,q) &= \frac{1}{2}\big(k_j+q_j)\G{j}\g{0}\G{\mu} + \frac{1}{2}k_j\G{\mu}\g{0}\G{j}. \label{eq:Vertexeps}
\ea
\end{subequations}
From the vertices and \eqqref{eq:Pigeneric1} it follows that $\Pi^{\mu\nu}_{\veps e}(q) = \Pi^{\nu\mu}_{e \veps}(-q)$, and similarly for other mixed current-current correlators.

The response matrices from Eqs.\,\eqref{eq:Transport1a}, \eqref{eq:Transport1b} and \eqref{eq:Transport2} can now be derived by focussing on the antisymmetric part of the current-current response functions. Writing out the dependence on $\bq$ and $\bt$ explicitly, their antisymmetric part is given by $\Pi^k_{ab}(\w,\bq;\bt) \equiv \varepsilon^{ijk}\Pi^{ij}_{ab}(\w,\bq;\bt)/2$. This antisymmetric part is a vector itself that is spanned by\cite{Wurff2017} $\bt$ and $\bq$. We can therefore decompose the relevant current-current response functions as follows
\begin{subequations}
\ba
&i\Pi^k_{ee}(\w,\bq;\bt) = \scme q^k + \sahe \w t^k, \label{eq:DecompPiee} \\
&i\Pi^k_{\veps e}(\w,\bq;\bt) = \s_{\veps}^{\text{CME}} q^k + \alpha^{\text{ANE}}_{\veps}  T \w t^k, \label{eq:DecompPiepse} \\
&i\Pi^k_{\veps \veps}(\w,\bq;\bt) = C_{\veps\veps} q^k + \bar{\kappa}_{\veps}^{\text{THE}} T \w t^k,\label{eq:DecompPiepseps}\\
&i\Pi^k_{ep}(\w,\bq;\bt) =  \sigma_{\parallel}^{\text{CVE}}  q^k \nonumber\\
&\phantom{=}\qquad\qquad\qquad+ 2\big(\sigma_{\perp}^{\text{CVE}} - \sigma_{\parallel}^{\text{CVE}}\big)(\bq\cdot\hat{\bt})\hat{t}^k,\label{eq:DecompPiep} \\
&i\Pi^k_{\veps p}(\w,\bq;\bt) = \sigma_{\veps,\parallel}^{\text{CVE}}  q^k\nonumber\\
&\phantom{=}\qquad\qquad\qquad+ 2\big(\sigma_{\veps,\perp}^{\text{CVE}} - \sigma_{\veps, \parallel}^{\text{CVE}}\big)(\bq\cdot\hat{\bt})\hat{t}^k, \label{eq:DecompPiepsp}
\ea 
\end{subequations}
where all transport coefficients are a function of $\w$ and $\bq$. A few remarks are in order about these decompositions:
\begin{itemize}
\item[1)] In principle there could be terms proportional to $(\bq\times\bt)^k$ present in the decompositions as well. For $a,b\in\{e,\veps\}$ we have $\Pi^k_{ab}(\w,\bq;\bt) = - \Pi^k_{ab}(\w,-\bq;-\bt)$, meaning that such terms are not allowed. The other two correlators do not obey this symmetry and such terms are thus allowed. However, they can be shown to vanish in the long-wavelength limit and we therefore do not consider them here.
\item[2)] The last two decompositions do not contain a term proportional to $\w t^k$ like the first three because $i\Pi^k_{ep}(\w,{\bf0};\bt) = i\Pi^k_{\veps p}(\w,{\bf 0};\bt) = 0$.
\item[3)] The first two decompositions do not contain a term proportional to $(\bq\cdot\hat{\bt})\hat{t}^k$ because such a term would violate the residual invariance under time-independent gauge transformations that are consistent with our gauge choice $A_0 = 0$, which we use throughout this paper.
\item[4)] The term proportional to $C_{\veps\veps}$ is of no interest to us because it yields zero upon contracting it with $\eps_{ijk}q^jT/i\w$ to compute the energy current.
\end{itemize}
Using the decompositions in Eqs.\,\eqref{eq:DecompPiee}-\eqref{eq:DecompPiepsp}, the resulting current densities can easily be derived in linear response from \eqqref{eq:current}. As the magnetic field is given in momentum space by $B^i = i\varepsilon^{ijk}q^jA^k$ and the velocity can be written in terms of the vorticity as $v^i = \varepsilon^{ijk}\omega^j x^k$, we find, for instance, for the magnetovortical response in the electric current density
\ba
\langle J_e^i\rangle  &=  \Pi^{ij}_{ee}A^j + \Pi^{ij}_{ep}v^j \nonumber\\
&=\s^{\text{CME}}B^i + 2\sigma_{\parallel}^{\text{CVE}}\w^i \nonumber\\
&\phantom{=}- 2\varepsilon^{ijk}\big(\sigma^{\text{CVE}}_{\perp} - \sigma^{\text{CVE}}_{\parallel}\big)\hat{t}^l\hat{t}^k\partial^l v^j \nonumber\\
&=\s^{\text{CME}}B^i + 2\sigma^{\text{CVE}}_{\perp} \big(\delta^{ij} - \hat{t}^i\hat{t}^j\big)\w^j \nonumber\\
&\phantom{=}+ 2\sigma^{\text{CVE}}_{\parallel} \hat{t}^i\hat{t}^j\w^j.
\ea
Similarly, $\sigma^{\text{CVE}}_{Q,\perp}$ and $\sigma^{\text{CVE}}_{Q,\parallel}$ can be derived such that the full magnetovortical response matrix from \eqqref{eq:Transport1b} is obtained once all coefficients in the decompositions in Eqs.\,\eqref{eq:DecompPiee}, \eqref{eq:DecompPiep} and \eqref{eq:DecompPiepsp} have been calculated.

Using the same procedure we can derive the thermoelectric response matric in \eqqref{eq:Transport2} from the decompositions in Eqs.\,\eqref{eq:DecompPiee}, \eqref{eq:DecompPiepse} and \eqref{eq:DecompPiepseps}. However, as explained in the previous section, the coefficients $\alpha_{\veps}^{\text{ANE}}$ and $\bar{\kappa}_{\veps}^{\text{THE}}$ do not yet constitute the actual anomalous Nernst and thermal Hall transport coefficients. To obtain $\alpha^{\text{ANE}}$ and $\bar{\kappa}^{\text{THE}}$ we need to consider the response of the \emph{heat} current and not the energy current. Besides that we also need to subtract the superfluous contribution coming from the rotating currents due to the electric and heat orbital magnetizations $M^{\text{orb}}_e$ and $M^{\text{orb}}_Q$, respectively. We show in Sec.\ref{subsec:magncontrib} how these contributions naturally occur as diamagnetic-like contributions to the effective action and compute them explicitly.
%%%%%%%%%%%%%%%%%%%%%%%%%%%%%%%%%%%%%%%%%%%%%%%%%%%%%%%%%%%%%%%%%%%%%%%%%%%%%%%%%%%%%%%%%%%%%%%%%%%%%%%%%%%%%%%%%%%%%%%%%%%%%%%%%%%%%%%%%%%%%%%%%%%%%%%%%%%%%%%%%%%%%%%%%%%%%%%%%%%%%%%%%%%%%%%%%%%%%%%%%%%%%%%%%%%%%%%%%%%%%%%%%%%%%%%%%%%%%%%%%%%%%%%%%%%%%%%%%%%%%%%%%%%%%%%%%%%%%%%%%%%%%%%%%%%%%%%%%%%%%%%%%%%%%%%%%%%%%%%%%%%%%%%%%%%%%%%%%%%%%%%%%%%%%%%%%%%%%%%%%%%%%%%%%%%%%%%%%%%%%%%%%%%%%%%%%%%%%%%%%%%%%%%%%%%%%%%%%%%%%%%%%%%%%%%%%%%%%%%%%%%%%%%%%%%%%%%%%%%%%%%%%%%%%%%%%%%%%%%%%%%%%%%%%%%%%%%%%%%%%%%%%%%%%%%%%%%%%%%%%%%%%%%%%%%%%%%%%%%%%%%%%%%%%%%%%%%%%%%%%%%%%%%%%%%%%%%%%%%%%%%%%%%%%%%%%%%%%%%%%%%%%%%%%%%%%%%%%%%%%%%%%%%%%%%%%%%%%%%%%%%%%%%%%
\subsection{Explicit expressions for the current-current response functions} \label{subsec:currentcurrentexplicit}
As an example, let us consider the computation of $i\Pi_{e p}^{k}(\w,\bq;\bt)$ in some more detail. All other current-current correlators can be computed in a similar manner, albeit in terms of lengthier expressions. Upon restricting to spatial indices and going from real to imaginary time, we find, using \eqqref{eq:Pigeneric1} and the vertices in \eqqref{eq:Vertexe} and \eqqref{eq:Vertexp},
\ba
&i\Pi^{ij}_{ep}(i\w_b,\bq;\bt) = -\frac{ie}{2\beta}\sum_{\chi,i\w_n}\chi\int_{\bk}\,(2k^{j} + q^{j}) \nonumber\\
&\times\text{Tr}\big[(\s^{i} + \chi t^{i})G_{\chi}(i\w_n + i\w_b, \bk+\bq)G_{\chi}(i\w_n,\bk)\big] \label{eq:Piep1}, 
\ea
with $i\w_b$ an external bosonic Matsubara frequency and $\int_{\bk} \equiv \int \text{d}^3\bk/(2\pi)^3$. Substituting the Matsubara Green's function from \eqqref{eq:Green}, we perform the trace and Matsubara sum\cite{Stoof2009} and Wick rotate back to real frequencies to find for the retarded current-current correlator
\ba 
\hspace{-.1cm}i\Pi^{ij}_{ep}(\w,\bq;\bt) &=-\frac{ie}{4}\!\sum_{\chi,u,v=\pm}\!u \int_{\bk} \frac{N^{\chi}_{uv}(\w,\bq,\bt,\bk)}{|\bk+\bq|} \nonumber\\
&\phantom{=}\times\big[ iv\chi(\bq\times\hat{\bk})^i + uv|\bk+\bq| \hat{k}^i + k^i \nonumber\\
&\phantom{=}\quad\,\, + q^i + v(|\bk| + \hat{\bk}\cdot\bq)t^i \big](2k^j + q^j) \label{eq:Piep2},
\ea
where we defined the function\cite{Warringa2009}
\ba
&N^{\chi}_{uv}(\w,\bq,\bt,\bk) \equiv \frac{\NFD{\veps_{v\bk} - \mux} - \NFD{\veps_{u\bk+\bq} - \mux}}{\w^+ +\veps_{v\bk} -  \veps_{u\bk+\bq}} \nonumber \\
&=\frac{v\NFD{v\veps_{v\bk} - v\mux} - u\NFD{u\veps_{u\bk+\bq} - u\mux} +\frac{1}{2}(u-v)}{\w^+ +\veps_{v\bk} -  \veps_{u\bk+\bq}}, \nonumber
\ea
with $\NFD{x}\equiv (e^{\beta x}+1)^{-1}$ the Fermi-Dirac distribution and $\veps_{n\bk}$ the dispersion relation defined in \eqqref{eq:energy}. In the second step we used the identity $\NFD{x} = 1 - \NFD{-x}$, such that the terms proportional to $(u-v)/2$ explicitly represent the contribution from the Dirac sea that yields an ultraviolet divergence when integrated in \eqqref{eq:Piep2} over all $\bk$. For the first term in \eqqref{eq:Piep2}, which is proportional to $\chi$, this divergence is exactly cancelled upon carrying out the sum over the two cones. In fact, the term proportional to $\chi$ is the only one of interest to us, because the other terms in \eqqref{eq:Piep2} do not contribute to the antisymmetric part of the current-current correlation function in the long-wavelength limit. This can be shown by realizing that
\ba
N^{\chi}_{uv}(\w,\bq,\bt,\bk) &= N^{\chi}_{uv}(\w,-\bq,-\bt,-\bk) \nonumber\\
&= \big[N^{\chi}_{vu}(-\w,-\bq,\bt,\bk+\bq)\big]^* \label{eq:N2}. 
\ea
Using these relations to explicitly symmetrize the integrand in \eqqref{eq:Piep2} and subsequently going to the long-wavelength limit shows that the last four terms do not contribute in this limit. Therefore we focus on the antisymmetric part of the first term in \eqqref{eq:Piep2}. In the end, we can write the remaining result as 
\be
i\Pi_{ab}^{k}(\w,\bq;\bt) = \frac{1}{4}\sum_{\chi,u,v = \pm}\!\chi\! \int_{\bk} N^{\chi}_{uv}(\w,\bq,\bt,\bk) f^{k,uv}_{ab}(\bq,\bt,\bk), \label{eq:Pigeneric2}
\ee
with 
\be
f^{k,uv}_{ep}(\bq,\bt,\bk) =\frac{euv(2\bk + \bq)\cdot\big(\bq\,k^k - \bk\,q^k\big)}{2|\bk||\bk+\bq|}. \label{eq:fep}
\ee
In a similar fashion we obtain $f_{ab}^{k,uv}(\bq,\bt,\bk)$ for the other current-current correlators. As they are rather lengthy, we present them in Appendix \ref{app:explicitexpressions}. The most important point to note is that $f_{ee}^{k,uv}(\bq,\bt,\bk)$, $f_{e\veps}^{k,uv}(\bq,\bt,\bk)$ and $f^{k,uv}_{\veps e}(\bq,\bt,\bk)$ contain terms that do not vanish at $\bq = {\bf 0}$. It is exactly these terms that lead to $\sigma^{\text{AHE}}$, $\alpha_{\veps}^{\text{ANE}}$ and $\bar{\kappa}^{\text{THE}}_{\veps}$ in the decompositions in Eqs.\,\eqref{eq:DecompPiee}, \eqref{eq:DecompPiepse} and \eqref{eq:DecompPiepseps}. 

Now that we have explained how to obtain and decompose the relevant current-current correlation function, we use them in the next section to calculate the magnetovortical response in several cases of interest. 
%%%%%%%%%%%%%%%%%%%%%%%%%%%%%%%%%%%%%%%%%%%%%%%%%%%%%%%%%%%%%%%%%%%%%%%%%%%%%%%%%%%%%%%%%%%%%%%%%%%%%%%%%%%%%%%%%%%%%%%%%%%%%%%%%%%%%%%%%%%%%%%%%%%%%%%%%%%%%%%%%%%%%%%%%%%%%%%%%%%%%%%%%%%%%%%%%%%%%%%%%%%%%%%%%%%%%%%%%%%%%%%%%%%%%%%%%%%%%%%%%%%%%%%%%%%%%%%%%%%%%%%%%%%%%%%%%%%%%%%%%%%%%%%%%%%%%
%%%%%%%%%%%%%%%%%%%%%%%%%%%%%%%%%%%%%%%%%%%%%%%%%%%%%%%%%%%%%%%%%%%%%%%%%%%%%%%%%%%%%%%%%%%%%%%%%%%%%%%%%%%%%%%%%%%%%%%%%%%%%%%%%%%%%%%%%%%%%%%%%%%%%%%%%%%%%%%%%%%%%%%%%%%%%%%%%%%%%%%%%%%%%%%%%%%%%%%%%%%%%%%%%%%%%%%%%%%%%%%%%%%%%%%%%%%%%%%%%%%%%%%%%%%%%%%%%%%%%%%%%%%%%%%%%%%%%%%%%%%%%%%%%%
\section{Anisotropic magnetovortical transport}  \label{sec:chiral}
In order to calculate the magnetovortical effects, we thus need to calculate the contributions in Eqs.\,\eqref{eq:DecompPiee}-\eqref{eq:DecompPiepsp} that are linear in $q^k$. From, e.g., \eqqref{eq:fep} we see that the current-current response functions contain terms in their integrands that are already explicitly proportional to $q^k$, but also terms proportional to $k^k$. The latter types of term give, upon integration, terms proportional to $q^k$ and $t^k$. By forming linear combinations of the projections of $\Pi^k_{ab}(\w,\bq;\bt)$ onto $q^k$ and $t^k$ we can extract the part that is proportional to $q^k$. For instance,
\ba
\sigma^{\text{CME}}(\w,\bq) = \bigg[\frac{|\bt|^2q_k - (\bq\cdot\bt)t_k}{|\bq|^2|\bt|^2 - (\bq\cdot\bt)^2}\bigg]i\Pi_{ee}^k(\w,\bq;\bt), \label{eq:cme}
\ea
and similarly for the other transport coefficients. We will calculate the magnetovortical transport coefficients both in the long-wavelength limit and compute their frequency dependence, starting with the former.

%%%%%%%%%%%%%%%%%%%%%%%%%%%%%%%%%%%%
%%%%%%%%%%%%%%%%%%%%%%%%%%%%%%%%%%%%
%%%%%%%%%%%%%%%%%%%%%%%%%%%%%%%%%%%%%%%%%%%%%%%%%%%%%%%%%%%%%%%%%%%%%%%%%%%%%%%%%%%%%%%%%%%%%%%%%%%%%%%%%%%%%%%%%%%%%%%%%%%%%%
\subsection{Long-wavelength limit} \label{subsec:lwl}
It is well-known that the long-wavelength limit of current-current response functions depends on the order of limits\cite{Mahan2013}. Therefore, we expand the expressions for the current-current response functions for small $|\bq|$ and $\w$, but keep the fraction $x\equiv \w/v_F|\bq|$ fixed, such that $x\rightarrow0$ corresponds to the static limit and $x\rightarrow\infty$ to the homogeneous, or transport limit. Using this procedure on $N_{uv}^{\chi}(\w,\bq,\bt,\bk)$ yields a Fermi sea (interband) contribution when $uv = -1$, i.e., at $T=0$ we find
\ba 
&N^{\chi}_{-+}(\w,\bq,\bt,\bk) \nonumber\\
&\phantom{=}\approx \bigg[1-\frac{\w}{2v_F|\bk|}+\frac{\bq\cdot\bt}{2|\bk|}-\frac{\bk\cdot\bq}{2|\bk|^2}\bigg]\frac{\NFD{\veps_{+\bk}-\mux}}{2v_F|\bk|} ,\label{eq:Nplusmin}
\ea 
from which $N_{+-}(\w,\bq,\bt,\bk)$ follows by using \eqqref{eq:N2}. When $u=v = +1$ we find a Fermi surface (intraband) contribution given by
\ba 
\!\!\!\!\!N^{\chi}_{++}(\w,\bq,\bt,\bk) \overset{|\bq|\rightarrow0}{=} -\frac{\big(\hat{\bq}\cdot\bt + \hat{\bk}\cdot\hat{\bq}\big)N_{\text{F}}'(\veps_{+\bk}-\mux)}{\w/v_F|\bq| - \hat{\bq}\cdot\bt-\hat{\bk}\cdot\hat{\bq}}, \!\!\label{eq:FermiSurface}
\ea
with $N'_{\text{F}}(x)$ the derivative of the Fermi-Dirac distribution. We thus observe that in the homogeneous limit this last term does not contribute, whereas in the static limit it does contribute. Finally, we note that at $T=0$ the term with $u,v=-$ yields zero. 
%%%%%%%%%%%%%%%%%%%%%%%%%%%%%%%%%%%%%%%%%%%%%%%%%%%%%%%%%%%%%%%%%%%%%%%%%%%%%%%%%%%%%%%%%%%%%%%%%%%%%%%%%%%%%%%%%%%%%%%%%%%%%%%%%%%%%%%%%%%%%%%%%%%%%%%%%%%%%%%%%%%%%%%%%%%%%%%%%%%%%%%%%%%%%%%%%%%%%%%%%%%%%%%%%%%%%%%%%%%%%%%%%%%%%%%%%%%%%%%%%%%%%%%%%%%%%%%%%%%%%%%%%%%%%%%%%%%%%%%%%
\begin{figure}[t!]
\includegraphics[scale=.85]{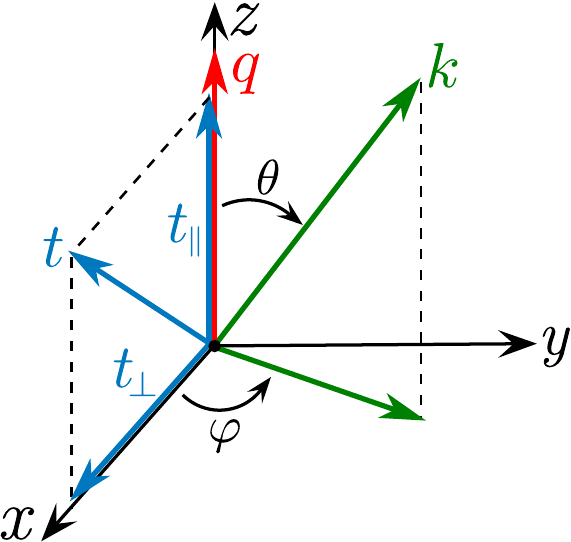}
\caption{Coordinate system used to calculate the tilt-dependent magnetovortical transport coefficients in the long-wavelength limit. We choose the coordinate system such that the vector $\bq$ (depicted in red) lies along the $z$-axis and the vector $\bt$ (depicted in blue) in the $xz$-plane. Therefore: $\bk\cdot\bq = |\bk||\bq|\cos\theta$, $\bq\cdot\bt=q \tpara$ and $\bk\cdot\bt = k (t_{\perp} \cos\varphi\sin\theta + \tpara \cos\theta)$ with $\varphi \in [0,2\pi)$ and $\theta\in[0,\pi]$.} \label{fig:CoordinateSystem}
\end{figure}
%%%%%%%%%%%%%%%%%%%%%%%%%%%%%%%%%%%%%%%%%%%%%%%%%%%%%%%%%%%%%%%%%%%%%%%%%%%%%%%%%%%%%%%%%%%%%%%%%%%%%%%%%%%%%%%%%%%%%%%%%%%%%%%%%%%%%%%%%%%%%%%%%%%%%%%%%%%%%%%%%%%%%%%%%%%%%%%%%%%%%%%%%%%%%%%%%%%%%%%%%%%%%%%%%%%%%%%%%%%%%%%%%%%%%%%%

\begin{figure}[t!]
\includegraphics[scale=.5]{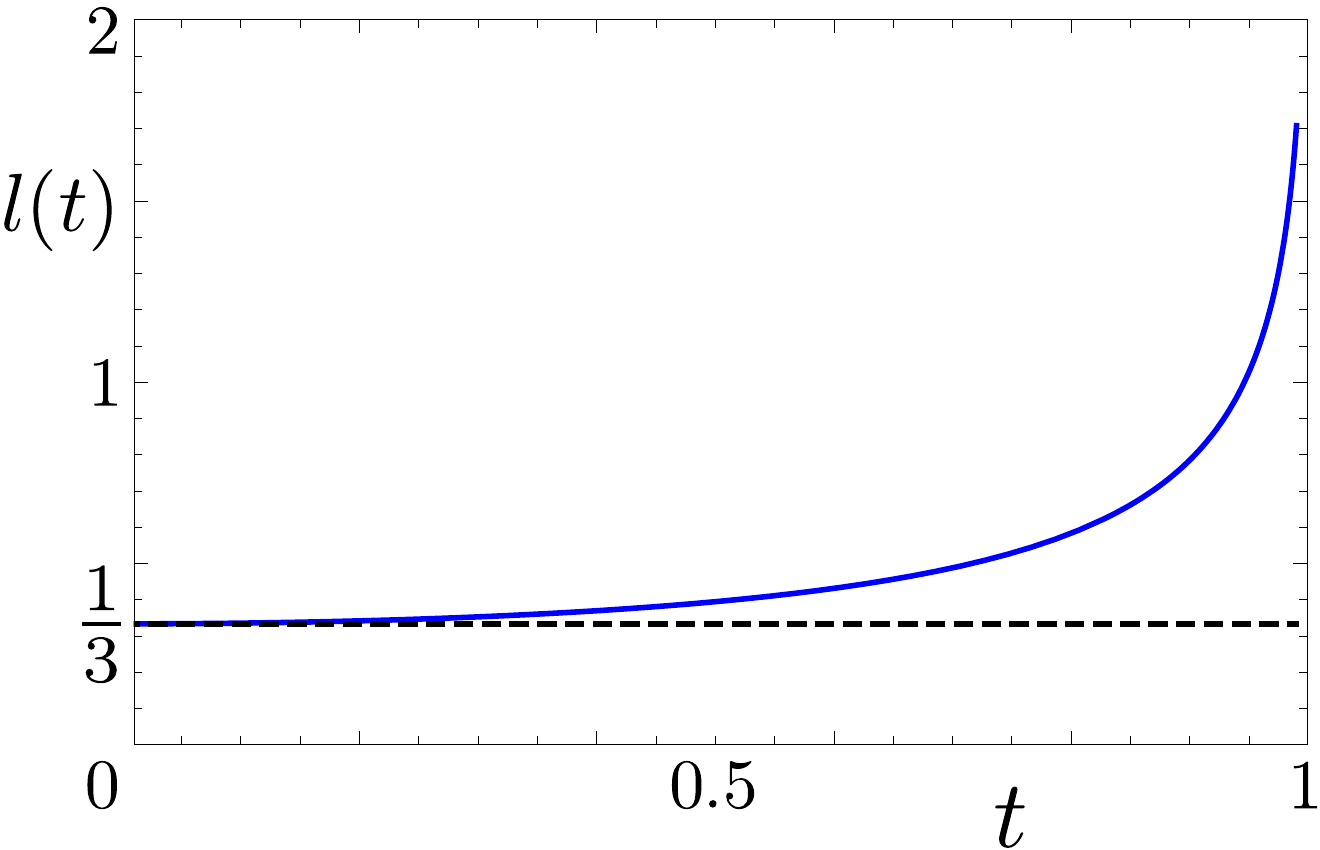}
\caption{Plot of the function $l(t)$ from \eqqref{eq:l}. It goes to the constant value of $1/3$ for small $t$ (indicated by the dashed black line) and diverges when $t\rightarrow1$, which signals the Lifshitz transition from a type-I to a type-II Weyl cone.} \label{fig:l}
\end{figure}

\begin{figure}[b!]
\includegraphics[scale=.6]{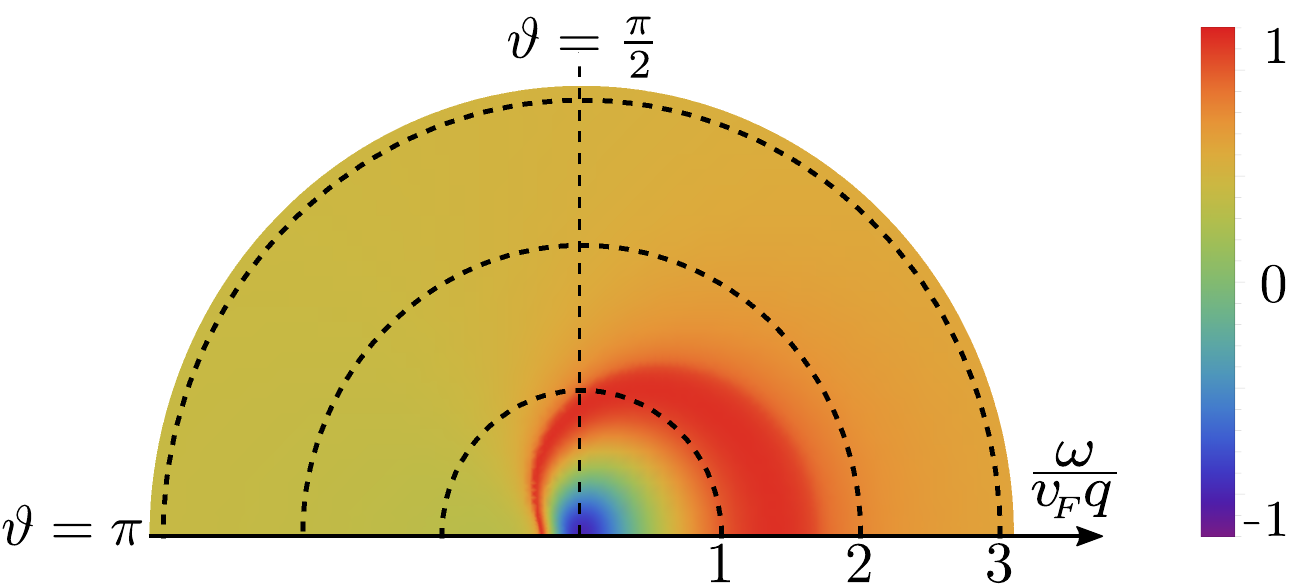}
\caption{Polar plot of the function $W(x,\bt)$ from \eqqref{eq:W} for $t=7/10$. The radial coordinate is $x = \w/v_F q$ and the angle $\vartheta$ is defined by $\bq\cdot\bt = |\bq||\bt|\cos\vartheta$. The homogeneous limit is obtained for large radii and becomes angle-independent. Likewise, the static limit is reached for vanishing radii and also becomes angle-independent.} \label{fig:Wpolar}
\end{figure}

Armed with these expansions we calculate the magnetovortical response in the long-wavelength limit. Using the coordinate system illustrated in \figgref{fig:CoordinateSystem}, we find for the chiral magnetic conductivities
\begin{subequations}
\ba
\scme(\w/v_F|\bq|) &= \frac{e^2\mu_5}{2\pi^2}(1-t^2)W(\w/v_F|\bq|,\bt), \label{eq:scmelwl}\\
\s^{\text{CME}}_{\veps}(\w/v_F|\bq|) &= -\frac{e\mu\mu_5}{2\pi^2}\big[2(1-t^2)W(\w/v_F|\bq|,\bt) - 1\big] \label{eq:scmepslwl},
\ea
\end{subequations}
and similar, but lengthier, expressions for the vortical effects. The function $W(\w/v_F|\bq|,\bt)$ is given explicitly in Appendix \ref{app:explicitexpressions}. In general it depends on the angle between $\bq$ and $\bt$, but in the static and homogeneous limit it reduces to an angle-independent result given by
\be
\hspace{-.1cm}W(\w/v_F|\bq|,\bt) = \begin{dcases} (1-t^2)^{-1} \quad \!\!\text{for} \quad \w/v_F|\bq|\rightarrow0, \\ l(t) \qquad \!\!\qquad\text{for} \quad \w/v_F|\bq|\rightarrow\infty, \end{dcases} \label{eq:W}
\ee
in terms of the function, c.f. \figgref{fig:l},
\be
l(t)\equiv \frac{1}{2t^3}\log\bigg(\frac{1+t}{1-t}\bigg) - \frac{1}{t^2} \overset{t\rightarrow0}{=} \frac{1}{3}. \label{eq:l}
\ee
To illustrate its angle-dependence, we show a polar plot of the function $W(\w/v_F|\bq|;\bt)$ in Fig.\ref{fig:Wpolar}.
%%%%%%%%%%%%%%%%%%%%%%%%%%%%%%%%%%%%%%%%%%%%%%%%%%%%%%%%%%%%%%%%%%%%%%%%%%%%%%%
%%%%%%%%%%%%%%%%%%%%%%%%%%%%%%%%%%%%%%%%%%%%%%%%%%%%%%%%%%%%%%%%%%%%%%%%%%%%%%%%%%%%%%%%%%%%%%%%%%%%%%%%%%%%%%%%%%%%%%%%%%%%%%%%%%%%%%%%%%%%%%%%%%%%%%%%%%%%%%%%%%%%%%%%%%%%%%%%%%%%%%%%%%%%%%%%%%%%%%%%%%%%%%%%%%%%%%%%%
%%%%%%%%%%%%%%%%%%%%%%%%%%%%
%%%%%%%%%%%%%%%%%%%%%%%%%%%%%%%%%%%%%%%%%%%%%%%%%%%%%%%%%%%%%%%%%%%%%%%%%%%%%%%%%%%%%%%%%%%%%%%%%%%%%%%%%%%%%%%%%%%%%%%%%%%%%%%%%%%%%%%%%
%%%%%%%%%%%%%%%%%%%%%%%%%%%%%%%%%%%%%%%%%%%%%%%%%%%%%%%%%%%%%%%%%%%%%%%%%%%%%%%%%
\begin{figure*}[t!]
\includegraphics[scale=.49]{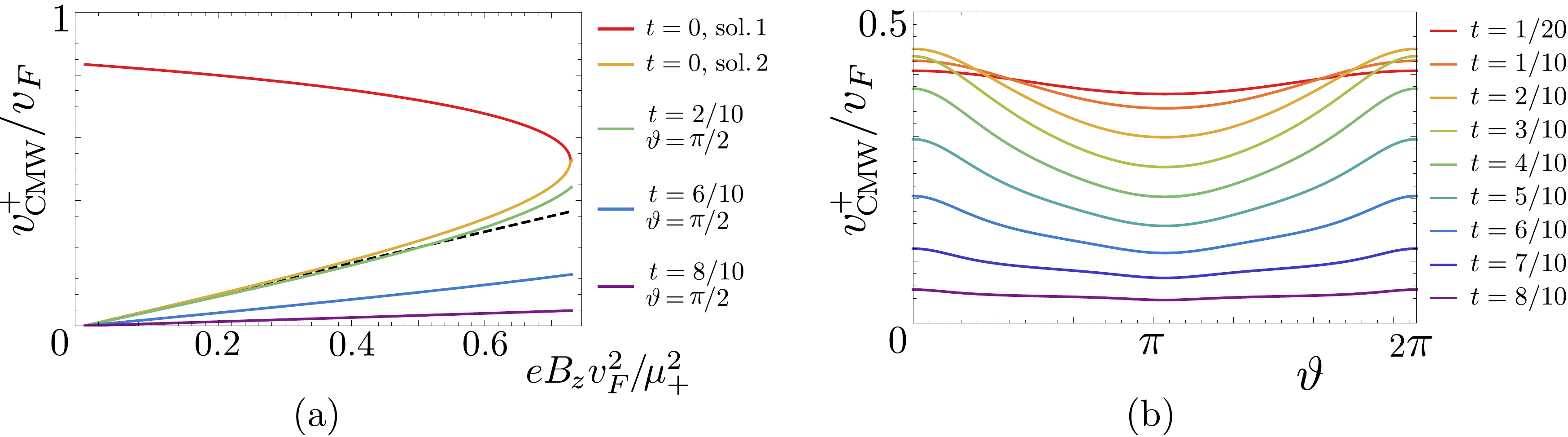}
\caption{(a) Plot of the chiral magnetic wave velocity (for a positive chirality) as a function of the dimensionless parameter $eB_z v_F^2/\mu_+^2$. The black dashed line indicates the result obtained by using the static-limit result for the susceptibility, which simply yields a linear function with a slope of $1/2$. The (from top to bottom) first and second curve give the highly damped and less-damped solutions at $t=0$, obtained by including the wavenumber and frequency dependence of the susceptibility in the long-wavelength limit. The last three plots show the least damped solution for several values of $t$ and ${\bf B}\cdot\bt = 0$. (b) Plot of the chiral magnetic wave velocity as a function of the angle $\vartheta$ between the magnetic field and the tilting direction. Here, we took $eB_zv_F^2/\mu_+^2 = 0.65$ and show the angle-dependence for several values of $t$. For small $t$, the velocity goes to a non-zero value and becomes angle-independent, whereas for $t\to 1$ its magnitude goes to zero.} \label{fig:CMEwave}
\end{figure*} %%%%%%%%%%%%%%%%%%%%%%%%%%%%%%%%%%%%%%%%%%%%%%%%%%%%%%%%%%%%%%%%%%%%%%%%%%%%%%%%%%%%%%%%%%%%%%%%%%%%%%%%%%%%%%%%%%%%

The fact that in the long-wavelength limit the transport properties depend on the tilting direction in a non-trivial way, is of importance for, for instance, the phenomenon of chiral magnetic waves \cite{Kharzeev2011,Star2015}. These are massless sound-like excitations in a fluid of chiral fermions. To understand how such excitations occur, we consider local fluctuations of the number densities $\delta n_{\pm}$ pertaining to the cone with chirality $\pm$. Assuming the fluctuations to be small, we may write $\delta \mu_{\pm} = \delta n_{\pm} / \chi_{\pm}$ in the current due to the chiral magnetic effect, with $\chi_{\pm} = \partial n_{\pm} / \partial\mu_{\pm}$ the corresponding susceptibilities. Considering the case of zero electric field, there is no chiral anomaly and we find from current conservation that
\be
\bigg(\partial_t \pm \frac{eB_z}{4\pi^2\chi_{\pm}}\partial_z\bigg) \delta n_{\pm} = 0, \label{eq:dispCMW}
\ee
where we took for simplicity a magnetic field in the $z$-direction. If the susceptibilities were constant and isotropic, the dispersion relations following from \eqqref{eq:dispCMW} would read $\w^{\pm}_{\bq}  = \pm v^{\pm}_{\text{CMW}}q_z$, with $v^{\pm}_{\text{CMW}} = e B_z / 4\pi^2\chi_{\pm}$. The susceptibilities, however, are anisotropic due to the tilting of the cones and in addition have a non-trivial frequency and wavenumber dependence. From the density-density response function $\Pi^{00}_{ee}(\w,\bq;\bt)/e^2$, we find for the susceptibilities
\ba
\hspace{-.2cm}\chi_{\pm}(&\w,\bq;\bt) = -\frac{1}{2}\sum_{u,v=\pm} \int_{\bk} \bigg(1 + uv\frac{|\bk|^2 +\bk\cdot\bq}{|\bk||\bk+\bq|}\bigg)\nonumber\\
&\phantom{=}\qquad\qquad\qquad\qquad\times N^{\pm}_{uv}(\w,\bq,\bt,\bk)\nonumber\\
&\phantom{===}\overset{|\bq|\to0}{=} \int_{\bk} \frac{(\hat{\bk}\cdot\hat{\bq} + \hat{\bq}\cdot\bt)N_{\text{F}}'(\veps_{+\bk}-\mu_{\pm})}{\w/v_F|\bq| - \hat{\bq}\cdot\bt-\hat{\bk}\cdot\hat{\bq}}, \label{eq:suscep}
\ea
where we used \eqqref{eq:FermiSurface} and reinstated $v_F$. The remaining integral can be performed exactly by using the coordinate system of \figgref{fig:CoordinateSystem} and yields an analytical result, c.f.\,Appendix \ref{app:explicitexpressions}, in terms of the function $W(x,\bt)$. At zero frequency the obtained expression reduces to
\be
\chi_{\pm}(0,{\bf 0};\bt)= \frac{\mu_{\pm}^2}{2\pi^2(1-t^2)^2v_F^3} = \frac{\partial n_{\pm}}{\partial \mu_{\pm}}, \label{eq:suscep2}
\ee
as it should. Fourier transforming \eqqref{eq:dispCMW} and using the result for $\chi_{\pm}(0,{\bf 0};{\bf 0})$ to make the resulting equation dimensionless, we find
\be
2\bigg(\frac{\w}{v_Fq_z}\bigg)\frac{\chi_{\pm}(\w,\bq;\bt)}{\chi_{\pm}(0,{\bf 0};{\bf 0})}= \pm \frac{eB_zv_F^2}{\mu_{\pm}^2}, \label{eq:selfconsistent}
\ee
from which the dispersion relation $\w_{\bq}$ can be found by solving the equation self-consistently. In our discussion we focus on the wave that propagates in the direction of the magnetic field, which corresponds to the plus sign in \eqqref{eq:selfconsistent}. To find the dispersion relation, we resort to numerics. For $t=0$, there are two solutions: one with a relatively high velocity that is highly damped and another solution with a lower velocity and corresponding lower damping. At a critical value for the dimensionless parameter $eB_zv_F^2/\mu_{+}^2$ these two solutions meet and above this value there are no solutions with a real part. We plot these solutions as a function of $eB_zv_F^2/\mu_{+}^2$ in \figgref{fig:CMEwave}(a), together with the result that is obtained by simply using the static-limit result for $\chi_{+}(\w,\bq;\bt)$. This plot shows that it is a rather good approximation to neglect the wavenumber dependence of the susceptibility, as only for relatively large values of $eB_zv_F^2/\mu_{+}^2$ the solutions start to deviate. In addition, we note that the expression for the susceptibility itself in \eqqref{eq:suscep} is only valid for weak magnetic fields. 
%%%%%%%%%%%%%%%%%%%%%%%%%%%%%%%%%%%%
\begin{table*}[!t]
\renewcommand\arraystretch{2.2}
\begin{tabular}{cccccccccc}
 & & \\
\multicolumn{1}{c}{} 
& \multicolumn{1}{c}{} 
& \multicolumn{1}{c}{} 
& \multicolumn{1}{c}{} 
& \multicolumn{2}{c|}{static limit} 
& \multicolumn{2}{c}{homogeneous limit} \\ \hline
\multicolumn{1}{c|}{conductivity} 
& \multicolumn{1}{c|}{type} 
& \multicolumn{1}{c|}{correlator} 
& \multicolumn{1}{c|}{units} 
& \multicolumn{1}{c|}{$t\neq0$} 
& \multicolumn{1}{c|}{$t=0$} 
& \multicolumn{1}{c|}{$t\neq0$} 
& \multicolumn{1}{c}{$t=0$}  \\ \hline \hline
\multicolumn{1}{c|}{$\mbox{\normalsize $\sigma^{\text{CME}}$ }$} 
& \multicolumn{1}{c|}{isotropic} 
& \multicolumn{1}{c|}{$\mbox{\normalsize $i\Pi^k_{ee} $ }$} 
& \multicolumn{1}{c|}{$e^2\mu_5/2\pi^2$} 
& \multicolumn{1}{c|}{$1$} 
& \multicolumn{1}{c|}{$1$} 
& \multicolumn{1}{c|}{$\displaystyle (1-t^2)l(t)$} 
& \multicolumn{1}{c}{$\displaystyle 1/3$}\\ \hline 
\multicolumn{1}{c|}{$\mbox{\normalsize $\sigma^{\text{CME}}_{\veps}$ }$} 
& \multicolumn{1}{c|}{isotropic} 
& \multicolumn{1}{c|}{$\mbox{\normalsize $i\Pi^k_{\veps e} $ }$} 
& \multicolumn{1}{c|}{$ e\mu \mu_5/2\pi^2$} 
& \multicolumn{1}{c|}{$\displaystyle -1$} 
& \multicolumn{1}{c|}{$\displaystyle -1$} 
& \multicolumn{1}{c|}{$\displaystyle 1-2(1-t^2)l(t)$} 
& \multicolumn{1}{c}{$\displaystyle 1/3$}\\ \hline 
\multicolumn{1}{c|}{$\mbox{\normalsize $\sigma^{\text{CME}}_{Q}$ }$} 
& \multicolumn{1}{c|}{isotropic} 
& \multicolumn{1}{c|}{$\mbox{\normalsize $i\Pi^k_{Q e} $ }$} 
& \multicolumn{1}{c|}{$e\mu \mu_5/2\pi^2$} 
& \multicolumn{1}{c|}{$\displaystyle 1$} 
& \multicolumn{1}{c|}{$\displaystyle 1$} 
& \multicolumn{1}{c|}{$\displaystyle 1$}  
& \multicolumn{1}{c}{$\displaystyle 1$}\\ \hline 
\multicolumn{1}{c|}{$\mbox{\normalsize $\sigma^{\text{CVE}}_{\parallel}$ }$} 
& \multicolumn{1}{c|}{longitudinal} 
& \multicolumn{1}{c|}{$\mbox{\normalsize $i\Pi^k_{e p} $ }$} 
& \multicolumn{1}{c|}{$e\mu\mu_5/2\pi^2$} 
& \multicolumn{1}{c|}{$\displaystyle -\frac{1}{(1-t^2)^2} $} 
& \multicolumn{1}{c|}{$\displaystyle -1$} 
& \multicolumn{1}{c|}{$\displaystyle -\frac{1}{2}\bigg(\frac{1}{1-t^2} - l(t)\bigg)$} 
& \multicolumn{1}{c}{$\displaystyle -1/3$} \\ \hline
\multicolumn{1}{c|}{$\mbox{\normalsize $\sigma^{\text{CVE}}_{\perp}$ }$} 
& \multicolumn{1}{c|}{transversal} 
& \multicolumn{1}{c|}{$\mbox{\normalsize $i\Pi^k_{e p} $ }$} 
& \multicolumn{1}{c|}{$ e\mu\mu_5/2\pi^2$} 
& \multicolumn{1}{c|}{$\displaystyle -\frac{2-t^2}{2(1-t^2)^2} $} 
& \multicolumn{1}{c|}{$\displaystyle -1$} 
& \multicolumn{1}{c|}{$\displaystyle -\frac{1}{4}\bigg(\frac{1}{1-t^2} + l(t)\bigg)$} 
& \multicolumn{1}{c}{$\displaystyle -1/3$} \\ \hline
\multicolumn{1}{c|}{$\mbox{\normalsize $\sigma^{\text{CVE}}_{\veps,\parallel}$ }$} 
& \multicolumn{1}{c|}{longitudinal} 
& \multicolumn{1}{c|}{$\mbox{\normalsize $i\Pi^k_{\veps p} $ }$} 
& \multicolumn{1}{c|}{$\mu_5(3\mu^2 + \mu_5^2)/6\pi^2$} 
& \multicolumn{1}{c|}{$\displaystyle \frac{1}{(1-t^2)^2}$} 
& \multicolumn{1}{c|}{$\displaystyle 1$} 
& \multicolumn{1}{c|}{$\displaystyle \frac{1}{4}\bigg(\frac{1-3t^2}{(1-t^2)^2} - 3l(t)\bigg)$} 
& \multicolumn{1}{c}{$\displaystyle 0$} \\ \hline
\multicolumn{1}{c|}{$\mbox{\normalsize $\sigma^{\text{CVE}}_{\veps,\perp}$ }$} 
& \multicolumn{1}{c|}{transversal} 
& \multicolumn{1}{c|}{$\mbox{\normalsize $i\Pi^k_{\veps p} $ }$} 
& \multicolumn{1}{c|}{$\mu_5(3\mu^2 + \mu_5^2)/6\pi^2$} 
& \multicolumn{1}{c|}{$\displaystyle \frac{2-t^2}{2(1-t^2)^2} $} 
& \multicolumn{1}{c|}{$\displaystyle 1$} 
& \multicolumn{1}{c|}{$\displaystyle -\frac{1}{8}\bigg(\frac{1+t^2}{(1-t^2)^2} - 3l(t)\bigg)$} 
& \multicolumn{1}{c}{$\displaystyle 0$} \\ \hline
\multicolumn{1}{c|}{$\mbox{\normalsize $\sigma^{\text{CVE}}_{Q,\parallel}$ }$} 
& \multicolumn{1}{c|}{longitudinal} 
& \multicolumn{1}{c|}{$\mbox{\normalsize $i\Pi^k_{Q p} $ }$} 
& \multicolumn{1}{c|}{$\mu_5(3\mu^2 + \mu_5^2)/6\pi^2$} 
& \multicolumn{1}{c|}{$\displaystyle -\frac{1}{2(1-t^2)^2}$} 
& \multicolumn{1}{c|}{$\displaystyle -1/2$} 
& \multicolumn{1}{c|}{$\displaystyle -\frac{1}{2(1-t^2)^2}$} 
& \multicolumn{1}{c}{$\displaystyle -1/2$} \\ \hline
\multicolumn{1}{c|}{$\mbox{\normalsize $\sigma^{\text{CVE}}_{Q,\perp}$ }$} 
& \multicolumn{1}{c|}{transversal} 
& \multicolumn{1}{c|}{$\mbox{\normalsize $i\Pi^k_{Q p} $ }$} 
& \multicolumn{1}{c|}{$\mu_5(3\mu^2 + \mu_5^2)/6\pi^2$} 
& \multicolumn{1}{c|}{$\displaystyle - \frac{2-t^2}{4(1-t^2)^2}$} 
& \multicolumn{1}{c|}{$\displaystyle-1/2$} 
& \multicolumn{1}{c|}{$\displaystyle - \frac{2-t^2}{4(1-t^2)^2}$} 
& \multicolumn{1}{c}{$\displaystyle -1/2$} \\ 
\end{tabular}
\caption{Table summarizing the results for the magnetovortical conductivities in the static and homogeneous limit. Note that the chiral magnetic conductivity has equal longitudinal and transverse parts in both limits, meaning that it is isotropic. All results presented here are at $T=0$ and for equal tilt of both cones. As all the tilt-dependend functions used here are invariant under $t\rightarrow -t$, these results hold for both inversion-symmetry breaking and inversion-symmetry retaining tilts.} \label{tab:magnetovortical}
\end{table*}
%%%%%%%%%%%%%%%%%%%%%%%%%%%%%%%%%%%%%%%%%%%%%%%%%%%%%%%%%%%%%%%%%%%%%%%%%%%%%%%%%%%%%%%%%%%%%%%%%%%%%%%%%%%%%%%%%%%%%%%%%%%%%%%%%%%%%%%%%%%%%%%%

The situation changes drastically upon tilting the cones. Firstly, the tilt renormalizes the magnitude of the susceptibility in \eqqref{eq:suscep}. As can be seen most clearly from the static-limit result in \eqqref{eq:suscep2}, the susceptibility becomes ever larger as $t\to1$, signaling the Lifschitz transition from a type-I to a type-II Weyl cone. From \eqqref{eq:selfconsistent} we see that this causes the velocity of the (least-damped) chiral wave to become significantly smaller as $t$ increases. We illustrate this behavior in \figgref{fig:CMEwave}(a). Another interesting consequence of tilting the cones is the fact that the dispersion relation becomes dependent on the angle $\vartheta$ between the magnetic field and the tilting direction. This is what we illustrate in \figgref{fig:CMEwave}(b) for one value of the dimensionless parameter $eB_zv_F^2/\mu_{+}^2$ and several values of $t$. From the figure we observe again that as $t$ grows, the velocity of the chiral magnetic wave goes to zero. Moreover, \figgref{fig:CMEwave}(b) shows that there is a quantitative difference between the case in which ${\bf B}$ is pointing in the same direction as $\bt$ ($\vartheta = 0$) and the case in which they are pointing in opposite directions ($\vartheta = \pi$). This is not surprising as the tilt breaks rotational invariance and introduces a preferred direction, which is also observable in \figgref{fig:Wpolar}.

To conclude this discussion of the anisotropic chiral magnetic wave, it is important to remark that in a material with a non-zero density of electrons, the dispersion of the chiral magnetic wave will inevitably be pushed up to the plasma frequency due to the fact that these chiral magnetic waves necessarily involve charge density fluctuations. A way around this is by considering not a single pair but rather two pairs of Weyl cones, yielding a total of four cones. We can then tune the chemical potentials such that the total chemical potential is zero, whereas the two pairs of cones have opposite chiral chemical potential. In this scenario, one of the chiral magnetic wave remains gapless because it does not involve fluctuations of the charge density.
%%%%%%%%%%%%%%%%%%%%%%%%%%%%%%%%%%%%%%%%%%%%%%%%%%%%%%%%%%%%%%%%%%%%%%%%%%%%%%%%%%%%%%%%%%%%%%%%%%%%%%%%%%%%%%%%%%%%%%%%%%%%%%%%%%%%%%%%%%%%%%%%%%%%%%%%%%%%%%%%%%%%%%%%%%%%%%%%%%%%%%%%%%%%%%%%%%%%%%%%%%%%%%%%%%%%%%%%%%%%%%%%%%%%%%%%%%%%%%%%%%%%%%%%%%%%%%%%%%%%%%%%%%%%%%%%%%%%%%%%%%%%%%%%%%%%%%%%%%%%%%%%%%%%%%%%%%%%%%%%%%%%%%%%%%%%%%%%%%%%%%%%%%%%%%%%%%%%%%%%%%%%%%%%%%%%%%%%%%%%%%%%%%%%%%%%%%%%%%%%%%%%%%%%%%%%%%%%%%%%%%%%%%%%%%%%%%%%%%%%%%%%%%%%
\subsection{Static and homogeneous limit} \label{subsec:stathomg}
Having studied the tilt-induced anisotropic behavior in the long-wavelength limit, we now specialize to two special cases of the long-wavelength limit: the static and the homogeneous limit. We summarize the results obtained for the various current-current correlators in the static and homogenous limit in Table \ref{tab:magnetovortical} and proceed by discussing the results presented in this table in detail. 

The first thing to note from Table \ref{tab:magnetovortical} is that the transport coefficients are non-zero in the static limit. In the case of the chiral magnetic conductivity this initially led to the believe that this constituted an equilibrium magnetic-field-driven current\cite{Burkov2011}, which appears to be unphysical as there can be no currents in equilibrium due to the Kohn theorem. The solution to this conundrum lies in the fact that the results in Table \ref{tab:magnetovortical} only hold for an energy difference $\Delta E = 0$ between the Weyl nodes\cite{Landsteiner2014B,Pesin2015}. Taking $\Delta E$ into account amounts to the replacement $\mu_5 \rightarrow\mu_5 + \Delta E/2$, leading to the vanishing of the currents in equilibrium since the system is in equilibrium when $2\mu_5 = -\Delta E$. 

Secondly, we note that our results for the vortical effects differ in the static and homogeneous limit from earlier obtained results for $t=0$ by Landsteiner \emph{et al.}\cite{Landsteiner2011,Landsteiner2013}. The reason is that the authors of these references use a different set of currents. We derived the momentum density in \eqqref{eq:T0mu} and energy current in \eqqref{eq:Tmu0} directly from the conservation laws that they obey and the equations of motion of the Dirac field. Landsteiner \emph{et al.}, instead, use the symmetrized energy-momentum tensor as the energy current. This coincides with $(J^i_p + J_{\veps}^i)/2$ in our definitions. Due to this symmetric definition they find that the chiral vortical conductivity $\s^{\text{CVE}}$ and chiral magnetic energy conductivity $\s^{\text{CME}}_{\veps}$ are the same in the static limit. In Table \ref{tab:Landsteiner} we show that upon taking the appropriate linear combinations, our results coincide at $t=0$ in both the static and the homogeneous limit with those of Landsteiner \emph{et al.}

Thirdly, it is interesting to note that the chiral magnetic conductivities are universal in the static limit, i.e., they do not dependent on the tilt of the Weyl cones. The other magnetovortical effects are, however, not universal in the static limit. The former can be understood by considering the Landau levels originating from a tilted Weyl cone. For a magnetic field in the $z$-direction, the dispersion relation of the chiral lowest Landau level is given by\cite{Goerbig2016}
\be
E_0^{\chi}(k_z) = \chi v_F\Big(t_z - \sqrt{1 - t_x^2-t_y^2}\Big)k_z,
\ee
under the assumption $t_x^2 + t_y^2 + t_z^2<1$. The lowest Landau level is thus still dispersing along the direction of the magnetic field, albeit with a renormalized slope. The higher Landau levels originating from the conduction and valence band each yield a zero net current and the Landau level degeneracy $eB/2\pi$ is not affected by the tilt. We can therefore obtain the charge current in the static limit from a one-dimensional integral along the $k_z$-direction, i.e.,
\ba
\!\!\langle \bJ_e \rangle &= -\frac{e^2{\bf B}}{2\pi}\!\! \sum_{\chi,u=\pm} \int_0^{\infty}\!\frac{\text{d}k_z}{2\pi} \frac{\text{d} E^{\chi}_0(k_z)}{\text{d} k_z } \NFD{E^{\chi}_0(k_z) - u\mux} \nonumber \\
&= \frac{e^2{\bf B}}{4\pi^2}\! \sum_{\chi,u = \pm}\!\chi u\! \int_0^{\infty}\text{d}\veps \, \NFD{\veps - u\mux} =  \frac{e^2\mu_5}{2\pi^2}{\bf B},\! \label{eq:CMEuniversal1}
\ea
which yields a universal answer because the density of states exactly cancels the slope of the lowest Landau level that determines the velocity. This also explains why the chiral magnetic energy current density is universal. Indeed, doing a similar calculation as in \eqqref{eq:CMEuniversal1}, we find
\ba
\langle \bJ_{\veps} \rangle &= -\frac{e{\bf B}}{4\pi^2}\!\! \sum_{\chi,u=\pm} \!\chi\! \int_0^{\infty}\!\!\text{d}\veps \,\veps \NFD{\veps - u\mux} \nonumber\\
&=-\frac{e\mu\mu_5}{2\pi^2}{\bf B}, \label{eq:CMEuniversal2}
\ea
which reproduces the corresponding result in Table \ref{tab:magnetovortical}. The overall minus sign in \eqqref{eq:CMEuniversal2} as compared to \eqqref{eq:CMEuniversal1} is due to the fact that the energy and charge current density differ by a factor of $-e$. The previous argument likewise clarifies why the chiral vortical conductivity $\sigma^{\text{CVE}}$ and the chiral vortical energy conductivity $\sigma_{\veps}^{\text{CVE}}$ depend in the same way on the tilt and only differ a factor of $-e$ in the static limit. 
%%%%%%%%%%%%%%%%%%%%%%%%%%%%%%%%%%%%%%%%%%%%%%%%%%%%%%%%%%%%%%%%%%%%%%%%%%%%%
%%%%%%%%%%%%%%%%%%%%%%%%%%%%%%%%%%%%%%%%%%%%%%%%%%%%
%%%%%%%%%%%%%%%%%%%%%%%%%%%%%%%%%%%%%%%%%%%%%%%%%%%%%%%%%%%%%%%%%%%%%%%%%%%%%%%%%%%%%%%%%%%%%%%%%%%%%%%%%%%%%%%%%%%%%%%%%%%%%%%%%%%%%%%%%%%%%%%%%%%%%%
\begin{table}[!t]
\normalsize
\centering
\begin{tabular}{c|c|c|c}
Result Ref.\cite{Landsteiner2014C} &units &stat.\,lim. &   hom.\,lim.  \\ \hline \hline
CME 
&$e^2\!\mu_5/2\pi^2$
&$ 1$    
&$ 1/3$ \\ \hline
CME-$\veps$ 
&$e\mu\mu_5/2\pi^2$
&$-1$    
&$0$ \\ \hline
CVE
&$e\mu\mu_5/2\pi^2$
&$-1$ 
&$0$  \\ \hline
CVE-$\veps$
&$\mu_5(3\mu^2\!+\!\mu_5^2)/2\pi^2$
&$1/3$           
&$0$                  
\end{tabular}
\caption{Table displaying the results from Landsteiner \emph{et al.} for the chiral magnetic, chiral vortical and chiral vortical energy conductivity in the static and homogeneous limit\cite{Landsteiner2014C}. We have $t=0$ and $T=0$ here. We reproduce these results by using the symmetrized energy-momentum tensor, resulting in the linear combination $i\big(\Pi^k_{e\veps} + \Pi^k_{ep}\big)/2$ for the chiral vortical conductivity and $i\big(\Pi^k_{\veps\veps} + 2\Pi^k_{\veps p}\big)/4$ for the chiral vortical energy conductivity. The latter does not contain a contribution from $\Pi^k_{pp}$ because its antisymmetric part vanishes. Note that with the symmetric definition of the energy-momentum tensor the chiral magnetic energy conductivity and chiral vortical conductivity necessarily coincide.} \label{tab:Landsteiner}
\end{table}
%%%%%%%%%%%%%%%%%%%%%%%%%%%%%%%%%%%%%%%%%%%%%%%%%%%%%%%%%%%%%%%%%%%%%%%%%%%%%%%%%%%%%%%%%%%%%%%%%%%%%%%%%%%%%%%%%%%%%%%%%%%%%%%%%%%%%%%%%%%%%%%%%%%%%%%%%%%%%%%%%%%%%%%%%%%%%%%%%%%%%%%%%%%%%%%%%%%%%%%%%%%%%%%%%%%%%%%%%%%%%%%%%%%%%%%%%%%%%%%%%%%%%%%%%%%%%%%%%%%%%%%%%%%%%%%%%%%%%%%%%%%%%%%%%%%%%%%%%%%%%%%%%%%%%%%%%%%%%%%%%%%%%%%%%%%%%%%%%%%%%%%%%%%%%%%%%%%%%%%

Another convenient framework to understand the universality of the chiral magnetic conductivities, as well as the non-universality and angle-dependence of the chiral vortical conductivity, is the semiclassical chiral kinetic theory \cite{Son2013,Stephanov2012,Stephanov2014}. In kinetic theory the semiclassical equation of motion for the velocity of a wavepacket in the band $n$ attains a correction in the direction of the magnetic field when the band has a non-zero Berry curvature \cite{Niu2010}. This so-called anomalous velocity results in the following simple expression for the chiral magnetic current density, i.e., 
\be
\!\!\!\langle {\bJ}_{e} \rangle = -e^2\!\!\sum_{n,\chi=\pm} \int_{\bk} \big[{\bm \Omega}_{n\chi}(\bk)\cdot \partial_{\bk}\veps_{n\bk}\big] N_{\text{F}}(\veps_{n\bk} - \mux) {\bf B}. \label{eq:CMEKinetic}
\ee
Here, the Berry curvature is given by ${\bf \Omega}_{n\chi}(\bk) \equiv {\bm \nabla}_{\bk}\times\langle u_{n\chi\bk} |i{\bm \nabla}_{\bk}|u_{n\chi\bk}\rangle = -n\chi\bk/2|\bk|^3$, in terms of the Bloch states $|u_{n\chi\bk}\rangle$ associated with \eqqref{eq:Ham}. Performing the integral in \eqqref{eq:CMEKinetic} then simply yields the universal result in \eqqref{eq:CMEuniversal1}. Note that the divergence due to the Dirac sea cancels because of the sum over chiralities. 

For the vortical conductivity a similar argument holds. By comparing the minimally-coupled Hamiltonian corresponding to \eqqref{eq:Ham}, i.e., $\mathcal{H}_{\chi}(\bk + e\bA)$, to the Hamiltonian $\mathcal{H}_{\chi}(\bk) - \bk\cdot{\bm v}$, it becomes clear that in the isotropic case the velocity ${\bm v}$ acts as an effective vector potential given by ${\bf A}_{\text{eff}} = -\veps_{n\bk} {\bm v} / e v_F^2$. Taking the rotation on both sides of this relation results in a vorticity that can be described by an effective magnetic field ${\bf B}_{\text{eff}} = -2\veps_{n\bk} {\bm \omega} / e v_F^2$. In the case of an isotropic single-particle energy, simply substituting this effective magnetic field into \eqqref{eq:CMEKinetic} yields
\ba
\hspace{-.2cm}\langle {\bJ}_{e} \rangle &= \frac{2e}{v_F^2}\!\sum_{n,\chi=\pm} \int_{\bk} \big[{\bm \Omega}_{n\chi}(\bk)\cdot \partial_{\bk}\veps_{n\bk}\big]\veps_{n\bk} N_{\text{F}}(\veps_{n\bk} - \mux)  {\bm \omega} \nonumber\\
&= -\frac{e\mu\mu_5}{2\pi^2}\frac{2{\bm \omega}}{v_F^2}, \label{eq:CVEKinetic}
\ea
where again the Dirac sea cancelled due to the sum over chiralities. The only question that remains to be answered is how the relation ${\bf A}_{\text{eff}} = -\veps_{n\bk} {\bm v} / e v_F^2$ changes when the dispersion relation is modified by a tilting of the cones. It is clear that in this case the effective gauge field and velocity can be decomposed into components pointing along and perpendicular to the tilting direction. This ultimately leads to a chiral vortical conductivity that has a longitudinal and transverse component, as can also be seen in Table \ref{tab:magnetovortical}. Unfortunately, we have not yet been able to find a simple argument for the appropriate effective magnetic field to reproduce the tilt-dependent longitudinal and transversal chiral vortical conductivities obtained from the Kubo formula.

Finally, we note that the results presented in Table \ref{tab:magnetovortical} have all been calculated at $T=0$ because the integrals otherwise cannot be performed exactly for non-zero tilt. In the case of zero tilt, the transport coefficients can be calculated exactly at $T\neq0$ in the static and homogeneous limit. The result is that $\scme$, $\s^{\text{CME}}_{\veps}$ and $\s^{\text{CVE}}_e$ do not change at non-zero temperature, whereas $\s^{\text{CVE}}_{\veps}$ attains an additional term proportional to $T^2$ that can be attributed to the mixed gauge-gravitional anomaly \cite{Landsteiner2011,Landsteiner2014B}. It stands to reason that similar behavior will be found when doing a numerical calculation at non-zero temperature that includes tilt of the cones.

%%%%%%%%%%%%%%%%%%%%%%%%%%%%%%%%%%%%%%%%%%%%%%%%%%%%%%%%%%%%%%%%%%%%%%%%%%%%%%%%%%%%%%%%%%%%%%%%%%%%%%%%%%%%
\subsection{AC response} \label{subsec:AC}
Having discussed the long-wavelength response in detail, we now turn our attention to the AC magnetovortical response. In order to obtain the non-zero-frequency response of the chiral magnetovortical effects, we extract the parts of the current-current correlators that are proportional to $q^k$ and subsequently evaluate the rest in the local limit $\bq = {\bf 0}$ while keeping $\w$ non-zero. In this limit the contribution from the Fermi surface (intraband) vanishes, as can be seen from \eqqref{eq:FermiSurface}, and only the Fermi sea (interband) contribution remains. Note that upon taking the zero-frequency limit in the AC conductivities we obtain in this section, the answers reduce to the homogeneous-limit results presented in Table \ref{tab:magnetovortical}. 

We start with the frequency dependence of the chiral magnetic effect. Using the procedure outlined above, we find 
\ba
\hspace{-.2cm}\sigma^{\text{CME}}(\w) &= -2e^2v_F^3\sum_{\chi = \pm}\chi \int_{\bk} \frac{\vartheta(\mux - \veps_{+\bk})}{(\w^+)^2 - 4v_F^2|\bk|^2}\nonumber\\
&\quad\times\!\bigg[1+ \hat{\bk}\cdot\bt  +4v_F^2 \frac{|\bk|^2-(\bk\cdot\hat{\bt})^2}{(\w^+)^2 - 4v_F^2|\bk|^2} \bigg]. \label{eq:PiCMEw}
\ea
The first term in this expression is proportional to $[(\w^+)^2-v_F^2|\bk|^2]^{-1}$, which can be split into two first-order poles. The second term, however, contains second-order poles. These appear due to the fact that the electric charge current-current correlator, as can be seen from \eqqref{eq:Pigeneric1}, is not automatically proportional to $q^k$ after performing the trace. To obtain the part that is linear in the external wavenumber, the propagator $S_F(k+q)$ has to be expanded in $q$, thereby yielding an additional propagator in the integrand. The chiral magnetic conductivity is thus obtained from the trace over three propagators, which is in fact a triangle diagram. This explains the occurence of double poles in \eqqref{eq:PiCMEw}. The chiral vortical conductivities, contrastingly, are already linear in $q^k$, as can be seen from e.g. \eqqref{eq:fep}, and therefore in this case no expansion of the propagator is necessary. The vortical conductivities therefore only contain single poles and are not due to a triangle diagram, but rather the more conventional bubble diagram.
%%%%%%%%%%%%%%%%%%%%%%%%%%%%%%%%%%%%%%%%%%%%%%%%%%%%%%%%%%%%%%%%%%%%%%%%%%%%%%%%%%%%%%%%%%%%%%%%%%%%
\begin{figure}[t!]
\includegraphics[scale=1.25]{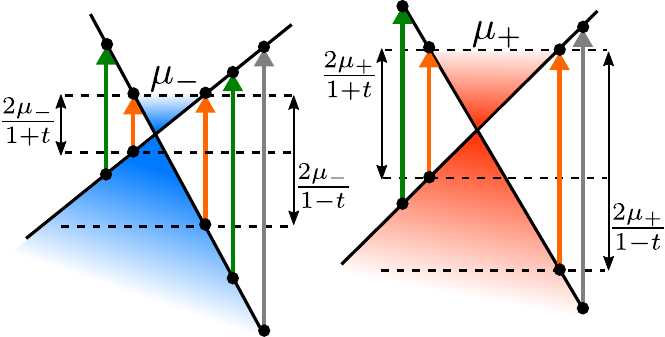}
\caption{Schematic representation of the optical absorption processes (i.e.\,with $\bq = {\bf 0}$) that are allowed in a pair of tilted Weyl cones (the tilt is chosen in the opposite direction of the momentum direction that defines the cones here). The orange arrows indicate the border of the frequency ranges defined by $\w_{\text{min/max}}$ and the green arrows transitions that are allowed within such a frequency range. The grey arrows indicate processes from deep in the Dirac sea that in principle are allowed, but destructively interfere upon subtracting the contributions from both cones.} \label{fig:OpticalTransitions}
\end{figure}
%%%%%%%%%%%%%%%%%%%%%%%%%%%%%%%%%%%%%%%%%%%%%%%%%%%%%%%%%%%%%%%%%%%%%%%%%%%%%%%%%%%%%%%%%%%%%%%%%%%%
%%%%%%%%%%%%%%%%%%%%%%%%%%%%%%%%%%%%%%%%%%%%%%%%%%%%%%%%%%%%%%%%%%%%%%%%%%%%%%%%%%%%%%%%%%%%%%%%%%%%%%%%%%%%%%%%%%%%
%%%%%%%%%%%%%%%%%%%%%%%%%%%%%%%%%%%%%%%%%%%%%%%%%%%%%%%%%%%%%%%%%%%%%%%%%%%%%%%%%%%%%%%%%%%%%%%%%%%%%%%%%%%%%%%%%%%%
\begin{figure*}
\includegraphics[scale=.73]{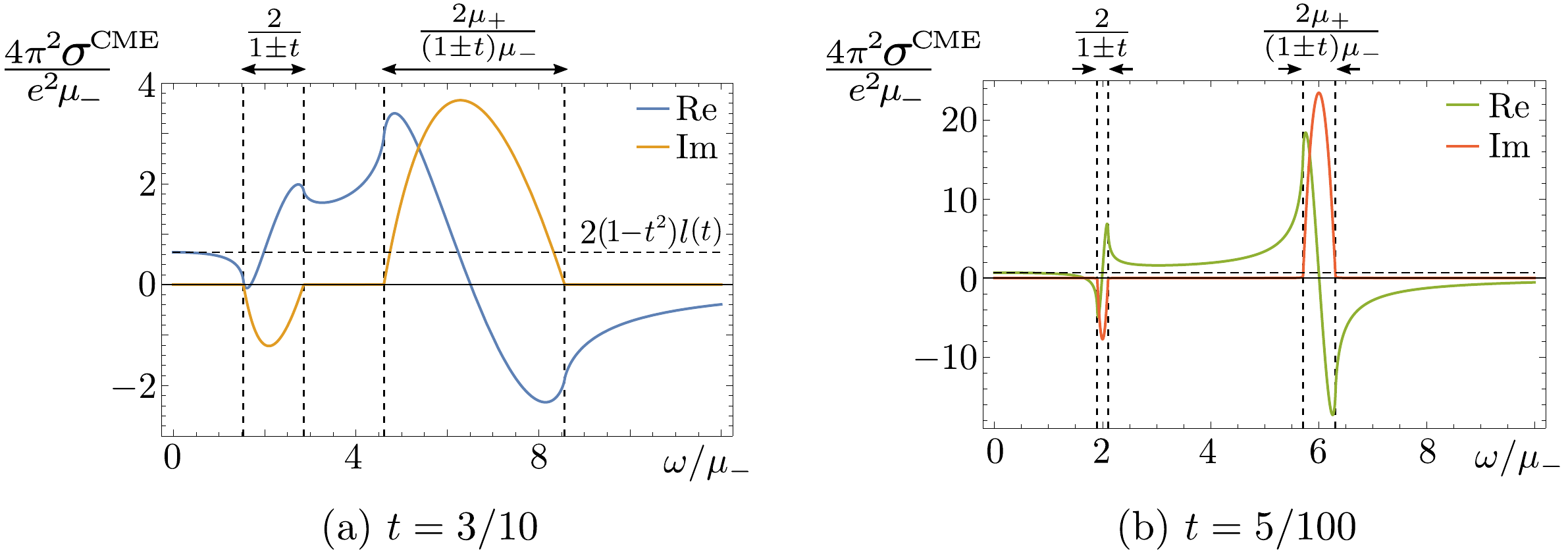}
\caption{Plots of the chiral magnetic conductivity from \eqqref{eq:CMCw} as a function of $\w/\mu_{-}$, normalized on $e^2\mu_-/4\pi^2$,  and for $\mu_+ / \mu_- = 3$. In (a) we used $t = 3/10$ to illustrate the tilt dependence and in (b) $t = 5/100$ to illustrate the convergence to the tilt-independent result from \eqqref{eq:CMEwt0Re} and \eqref{eq:CMEwt0Im}. In (a) it is clear that the imaginary part is only non-zero between $2\mux/(1\pm t)$, which is indicated by the vertical dashed lines in both figures. The limiting value for $\w/\mu_-\rightarrow0$ is given by $(\mu_+ / \mu_- - 1)(1-t^2)l(t)$ and is indicated by the horizontal dashed line in both figures.} \label{fig:freqdep1}
\end{figure*}
%%%%%%%%%%%%%%%%%%%%%%%%%%%%%%%%%%%%%%%%%%%%%%%%%%%%%%%%%%%%%%%%%%%%%%%%%%%%%%%%%%%%%%%%%%%%%%%%%%%%%%%%%%%%%%%%%%%%

Although the integral in \eqqref{eq:PiCMEw} can be performed analytically for non-zero tilt, it is illustrative to first consider the case of zero tilt, for which we find for the real and imaginary parts of the chiral magnetic conductivity\cite{Warringa2009,Landsteiner2014C}
\begin{subequations}\label{eq:CMEwt0}
\ba
\text{Re}\big[\s^{\text{CME}}(\w)\big] &=\sum_{\chi,u = \pm}\frac{\chi e^2\mux}{12\pi^2}\frac{\mux}{2\mux - u\w} \label{eq:CMEwt0Re} \\
\text{Im}\big[\s^{\text{CME}}(\w)\big] &=\sum_{\chi,u = \pm}\frac{\chi e^2\w^2 }{48\pi}u\delta(\w-2u\mux), \label{eq:CMEwt0Im}
\ea
\end{subequations}
which displays resonances around $\w = \pm2\mux$. Physically, this is due to the creation of an electron-hole pair by the excitation of a valence electron to the conduction band. Because of Pauli blocking this is for a single cone only possible when the externally applied frequency $\w$ obeys $|\w|>2\mu_+$. Upon subtracting the contributions from both cones, however, the transitions from deep in the Dirac sea, i.e., for $|\w|>2\mu_+$ when $\mu_+ > \mu_-$, destructively interfere. The fact that the imaginary part in \eqqref{eq:CMEwt0Im} contains delta functions, instead of the more conventional heaviside step functions, is precisely due to the fact that \eqqref{eq:PiCMEw} contains second-order poles, so that the answer is proportional to the derivative of these heaviside functions instead.

Tilting the cones yields four frequency intervals, rather than the four single frequencies $\w = \pm2\mux$, given by
\be
\wmin(\mux)\equiv\frac{2\mux}{1+t} < \w < \frac{2\mux}{1-t}\equiv \wmax(\mux), \label{eq:freqbands}
\ee
and similarly for negative $\omega$. In what follows we always use values for $\mu_{\pm}$ and $t$ such that $\w_{\text{min}}(\mu_+) > \w_{\max}(\mu_-)$. We thus expect the imaginary part of the chiral magnetic conductivity to be non-zero and finite within the two intervals defined by \eqqref{eq:freqbands}. Additionally, the real part should still show a resonance, albeit less pronounced than in \eqqref{eq:CMEwt0Re}. In \figgref{fig:OpticalTransitions} we present a graphical illustration of the allowed excitation processes for a pair of tilted cones.
%%%%%%%%%%%%%%%%%%%%%%%%%%%%%%%%%%%%%%%%%%%%%%%%%%%%%%%%%%%%%%%%%%%%%%%%%%%%%%%%%%%%%%%%%%%%%%%%%%%%%%%%%%%%%%%%%%%%
\begin{figure*}
\includegraphics[scale=.79]{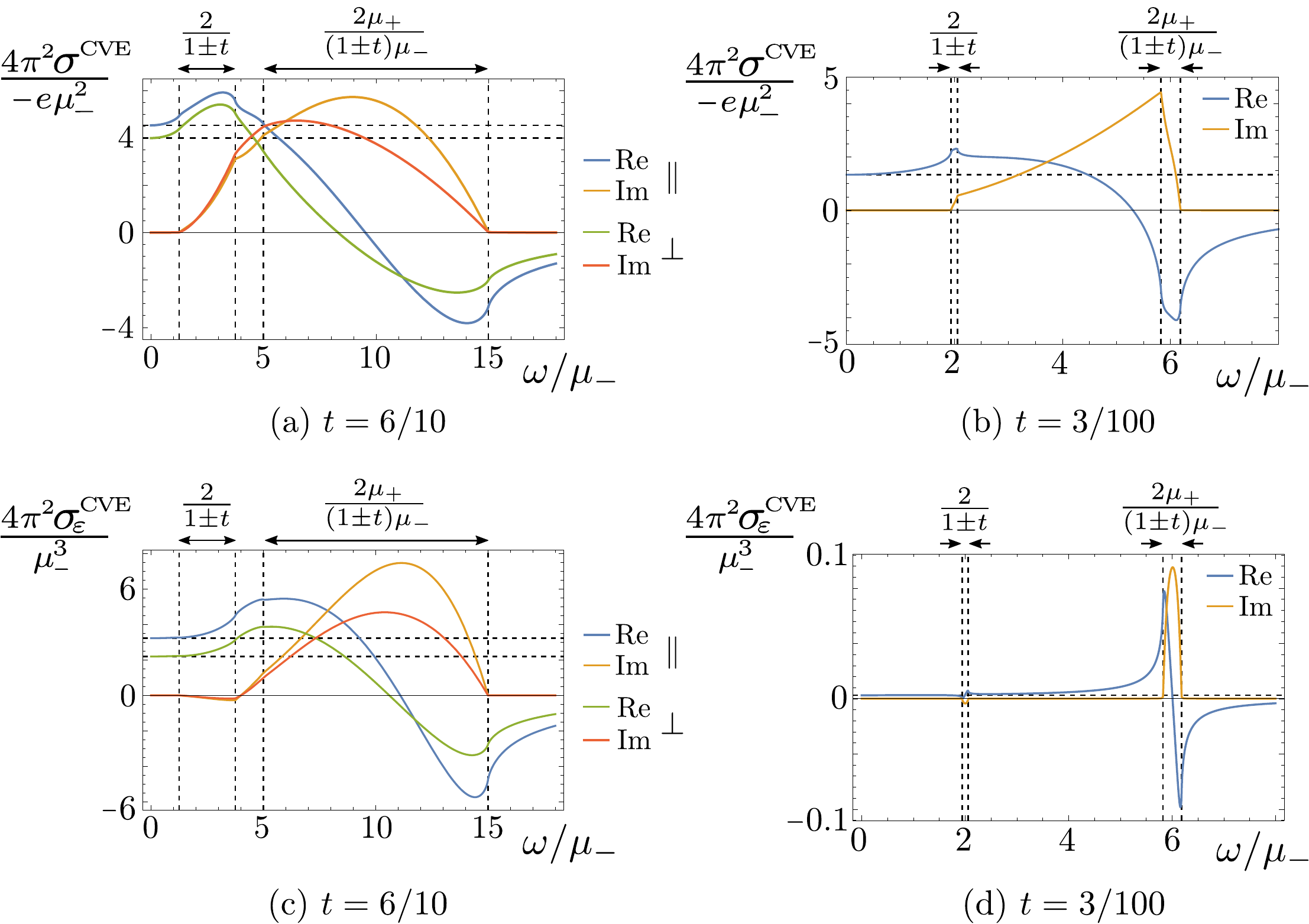}
\caption{In (a) and (b) we plot the anisotropic chiral vortical conductivity, normalized on $-e\mu_-^2/4\pi^2$ and in (c) and (d) the anisotropic chiral vortical energy conductivity, normalized on $\mu_-^3/4\pi^2$, both as a function of $\w/\mu_{-}$. In (a) we plot both $\sigma^{\text{CVE}}_{\perp}$ and $\sigma^{\text{CVE}}_{\parallel}$ for $t=6/10$ and $\mu_+ / \mu_- = 3$. The vertical dashed lines indicate the positions of $\w_{\text{min}}(\mu_{\pm})$ and $\w_{\text{max}}(\mu_{\pm})$. The horizontal dashed lines indicate the zero-frequency limiting values, given by the values listed in Table \ref{tab:magnetovortical}, multiplied by an additional factor of $(\mu_+^2/\mu_-^2 -1)$ for the chiral vortical conductivity and a factor of $(\mu_+^3/\mu_-^3 -1)$ for the chiral vortical energy conductivity, both of which are due to the normalization used in these plots. In (b) we plot only $\sigma^{\text{CVE}}_{\parallel}$ for $t=3/100$ because the difference with $\sigma^{\text{CVE}}_{\perp}$ is very small. In (c) and (d) we do the same for the chiral vortical energy conductivity.} \label{fig:freqdepCVE}
\end{figure*}
%%%%%%%%%%%%%%%%%%%%%%%%%%%%%%%%%%%%%%%%%%%%%%%%%%%%%%%%%%%%%%%%%%%%%%%%%%%%%%%%%%%%%%%%%%%%%%%%%%%%%%%%%%%%

We perform the integral in \eqqref{eq:PiCMEw} by choosing spherical coordinates $(\varphi,\theta,k)$ along the direction of $\bt$, such that $\bk\cdot\bt = kt\cos\theta$. This renders the integral over $\varphi$ trivial. The integral over $k$ can be written as an integral over the energy by employing the changes of variables $y = \cos\theta$ and $\veps = (1 +  t y)v_Fk$. The double integral can then be written as a product of an integral over $y$ and one over $\veps$. The former integral is easily performed, whereas the latter can be performed analytically only at $T=0$ for non-zero frequencies. Doing so, we find for the tilt and frequency-dependent chiral magnetic conductivity  
\ba
\hspace{-.2cm}\s^{\text{CME}}(\w) &= -\sum_{\chi=\pm}\frac{\chi e^2\mux (1-t^2)}{16\pi^2t^3}\bigg[2t -L_1(\w) \nonumber\\
&\phantom{=}\qquad\qquad+ \frac{(1-t^2)\w^2 + 4\mux^2}{4\w\mux}L_2(\w)\bigg], \label{eq:CMCw}
\ea
where we defined the functions\cite{Pesin2017}
\begin{subequations}
 \label{eq:Li}
\ba 
L_1(\w) &= \log\bigg(\frac{(\w^+)^2 - \wmax^2}{(\w^+)^2 - \wmin^2}\bigg), \label{eq:L1} \\
L_2(\w) &= \log\bigg(\frac{(\w^+ - \wmax)(\w^+ + \wmin)}{(\w^+ + \wmax)(\w^+ - \wmin)}\bigg),\label{eq:L2} \\
L_3(\w) &= \log\bigg(1-\frac{\wmin^2}{(\w^+)^2}\bigg) + \log\bigg(1-\frac{\wmax^2}{(\w^+)^2}\bigg) \label{eq:L3},
\ea
\end{subequations}
and omitted the dependence of $\w_{\text{min}/\text{max}}$ on the chemical potential for brevity. 

We plot the real and imaginary part of $\s^{\text{CME}}(\w)$ as a function of $\w/\mu_-$ in \figgref{fig:freqdep1}. For a tilt of $t = 3/10$ we observe the expected behavior in \figgref{fig:freqdep1}\,(a): in between $\wmin(\mu_{\pm})$ and $\wmax(\mu_{\pm})$ the chiral magnetic conductivity has an imaginary part, whereas the real part has a resonance that is broader and less steep than for the zero-tilt case. It is interesting to note that the imaginary part goes to zero exactly at $\wmin(\mu_{\pm})$ and $\wmax(\mu_{\pm})$ and is zero in between $\wmax(\mu_-)$ and $\wmin(\mu_+)$. The reason can be deduced from \eqqref{eq:CMCw}: both the function $L_1(\w)$ and $L_2(\w)$ contribute to the imaginary part, but exactly at $\wmin(\mux)$ and $\wmax(\mux)$ the function multiplying $L_2(\w)$ becomes equal to $1$, thereby cancelling the contribution from $L_1(\w)$, which simply comes with a factor of minus one. Hence, tilting the cone renders the chiral magnetic conductivity non-zero and finite, even around $\w = \pm2\mux$. In \figgref{fig:freqdep1}\,(b) we show that upon decreasing the tilt to $t = 3/100$, a very narrow resonance reappears. Clearly the results of \eqqref{eq:CMEwt0} are reproduced as $t$ goes to zero.

It is also interesting to note that a very similar calculation for the anomalous Hall effect shows that 
\be
\sigma^{\text{AHE}}(\w) = \frac{\sigma^{\text{CME}}(\w)}{v_F(1-t^2)}, \label{eq:cmeahe}
\ee
which only holds if we do not take the topological contribution due to the separation between the Weyl nodes into account. As the chiral magnetic conductivity remains non-zero when $t\rightarrow0$, the above equation seems to imply that the anomalous Hall conductivity also remains non-zero in this limit. However, the anomalous Hall current density does vanish because we defined it proportional to $t$: $\bJ^{\text{AHE}}(\w,\bq) = \s^{\text{AHE}}(\w,\bq)\bt\times {\bf E}$.

We now turn to the frequency dependence of the vortical effects. Following the same procedure, we find for instance for the electric charge current-momentum density correlator
\ba
\hspace{-.2cm}i\Pi^k_{ep}(\w,\bq;\bt) &\approx \frac{ev_F q^l}{4}\!\sum_{\chi = \pm}\chi\int_{\bk} \vartheta(\mux - \veps_{+\bk}) \big[\delta^{kl} - \hat{k}^k\hat{k}^l \big] \nonumber \\ 
&\phantom{=}\quad\times\bigg[\frac{1}{\w^+ - 2v_F|\bk|} - \frac{1}{\w^+ + 2v_F|\bk|}  \bigg]. \label{eq:Piepw}
\ea
From this expression we firstly observe one important difference with the chiral magnetic conductivity that we already alluded to before: there are no double poles present. We therefore expect a non-zero imaginary part in between $\wmax(\mu_-)$ and $\wmin(\mu_+)$. Furthermore, from \eqqref{eq:Piepw} it becomes clear that the vortical conductivity is anisotropic. The reason is that the integral over the term proportional to $\hat{k}^k\hat{k}^l$ can either yield a contribution proportional to $\delta^{kl}$, or a contribution proportional to $\hat{t}^k\hat{t}^l$, because these are the only symmetric tensors left after taking $|\bq|\to0$. 

The longitudinal and transverse vortical conductivities can also be expressed in terms of the functions $L_i(\w)$ defined in \eqqref{eq:Li} and we list the explicit, but lengthy expressions in Appendix \ref{app:freq}. Here, we instead plot the frequency dependence of the longitudinal and transversal chiral vortical conductivity and chiral vortical energy conductivity in \figgref{fig:freqdepCVE} for $t=6/10$ in (a) and (c) and for $t=3/100$ in (b) and (d). As expected, we observe a non-zero imaginary part on the whole frequency range defined by $\w_{\min}(\mu_-) < \w < \wmax(\mu_+)$. In addition, both the real and imaginary parts of the longitudinal conductivities are always larger than the transverse counterparts. This can be understood from the integrand in \eqqref{eq:Piepw}, which contains the transverse projection operator $\delta^{kl} - \hat{k}^k\hat{k}^l$. Because of the relative minus sign in this operator, the contribution of the term proportional to $\hat{k}^l\hat{k}^k$ to the transverse conductivity is negative. 

Furthermore, from \figgref{fig:freqdepCVE}(d) we note that in the limit of small tilt, the chiral vortical energy conductivity displays a resonance similar to the one for the chiral magnetic effect in \figgref{fig:freqdep1}(b). Its magnitude, however, is significantly smaller and upon taking the tilt to zero the chiral vortical energy conductivity vanishes. This is consistent with the result obtained in the homogeneous limit in Table \ref{tab:magnetovortical}. On the other hand, the small-tilt behavior of the chiral vortical conductivity in \figgref{fig:freqdepCVE}(b) is rather different. We observe that for small tilt it saturates to two peaks around $\w = 2\mu_{\pm}$, with a non-zero imaginary part in between. Upon inspecting the frequency behavior of the chiral magnetic energy conductivity, we find exactly the same behavior, but with an opposite sign. Adding both contributions, which is equivalent to using a symmetrized version of the energy-momentum tensor, thus yields zero for all frequencies, except at $\w = 0$. This result is consistent with previous work by Landsteiner \emph{et al.} \cite{Landsteiner2014C}. These authors use the symmetric energy-momentum tensor, a combination of Ward identities and rotational symmetry to show that only at $\w = 0$ there is a non-zero vortical response in the electric current. Physically, this means that all interband transitions are forbidden in their case. The situation in our case is different for two reasons: 1) we use a different set of currents and 2) rotational symmetry is broken by the tilting of the cones. We therefore have non-trivial frequency dependence for the magnetovortical conductivities. 

Before we turn our attention to the thermoelectric transport of tilted Weyl cones, let us summarize the main findings of this section. We started by calculating the long-wavelength response of the magnetovortical conductivities. The anisotropy introduced by the tilting of the cones led to an anisotropic velocity of the chiral magnetic waves. Subsequently we considered two specific cases of the long-wavelength limit: the static and the homogeneous limit, and listed all magnetovortical conductivities in Table \ref{tab:magnetovortical}. We found that the chiral magnetic conductivities remains isotropic, whereas the vortical conductivitities attain a transverse and longitudinal part. Moreover, the chiral magnetic conductivities turned out to be tilt-independent, or universal, in the static limit, which we managed to explain based on exact quantum and semiclassical arguments. Finally, we focussed on the AC magnetovortical response, finding rather different behavior for the vortical and magnetic conductivities. We explained how this ultimately is due to the fact that the chiral magnetic effect is determined by a triangle diagram, whereas the chiral vortical conductivities follow from a bubble diagram.
%%%%%%%%%%%%%%%%%%%%%%%%%%%%%%%%%%%%%%%%%%%%%%%%%%%%%%%%%%%%%%%%%%%%%%%%%%%%%%%%%%%%%%%%%%%%%%%%%%%%%%%%%%%%%%%%%%%%
%%%%%%%%%%%%%%%%%%%%%%%%%%%%%%%%%%%%%%%%%%%%%%%%%%%%%%%%%%%%%%%%%%%%%%%%%%%%%%%%%%%%%%%%%%%%%%%%%%%%%%%%%%%%%%%%%%%%
\section{Electronic and thermal transport} \label{sec:transport}
We now turn to the coupled off-diagonal thermoelectric transport from tilted Weyl cones as described by \eqqref{eq:Transport2}. Some of the results we discuss have already been presented by Bardarson \emph{et al.} \cite{Bardarson2017B}. These authors circumvent the need to subtract the superfluous contributions coming from unobservable, circulating currents of the form $\nabla \times {\bm M}_{e/Q}^{\text{orb}}$, with ${\bm M}^{\text{orb}}_e$ (${\bm M}_Q^{\text{orb}}$) the electric (heat) orbital magnetization density. In order to do this, they first calculate the anomalous Hall conductivity $\sahe$, which does not require any subtractions. Subsequently they use the Mott relation and Wiedemann-Franz law to calculate $\ane$ and $\kthe$ from $\sahe$.

Instead, we discuss here how these orbital magnetizations arise naturally as diamagnetic-like terms when performing linear response theory in the presence of a temperature gradient. We then calculate ${\bm M}^{\text{orb}}_e$ and ${\bm M}_Q^{\text{orb}}$ microscopically and explicitly subtract them from the currents coming from the Kubo formula to yield the transport currents. Finally, we discuss how the answers for the transport coefficients differ when considering inversion-symmetric or inversion-symmetry breaking tilt and discuss the consequences of a non-zero chiral chemical potential $\mu_5$. 

%%%%%%%%%%%%%%%%%%%%%%%%%%%%%%%%%%%%%%%%%%%%%%%%%%%%%%%%%%%%%%%%%%%%%%%%%%%%%%%%%%%%%%%%%%%%%%%%%%%%%%%%%%%%%%%%%%%%%%%%%%%%%%%%%%%%%%%%%%%%%%%%%%%%%%%%%%%%%%%%%%%%%%%%%%%%%%%%%%%%%%%%%%%%%%%%%%%%%%%%%%%%%%%%%%%%%%%%%%%%%%%%%%%%%%%%%%%%%%%
\subsection{Magnetization contributions} \label{subsec:magncontrib}
To see that magnetization contributions occur as diamagnetic-like terms when performing linear-response theory in the presence of a temperature gradient, consider an imaginary time action that contains a coupling between the electric current density $J_e^i(\bx,\tau)$ and an external vector potential $A^i(\bx,\tau)$, i.e.,
\be
S_{\text{coup.}} = \int\text{d}\bx\int_0^{\hbar\beta}\!\!\text{d}\tau A_i(\bx,\tau)J_e^i(\bx,\tau), \label{eq:magex1}
\ee
with $\beta = 1/k_BT$ in terms of the temperature $T$. Performing linear-response theory with this action shows that the first order contribution to the effective action for the gauge potential vanishes, because in equilibrium $\langle J^i_e(\bx,\tau)\rangle_0 = 0$. The contribution at second order, on the other hand, is non-zero and yields the current-current response functions we discussed in Sec.\ref{sec:currentcurrent}.

The situation changes if we now include temperature variations by writing $T(\bx) = T + \delta T(\bx)$. This causes the upper boundary of the integral over imaginary time in \eqqref{eq:magex1} to depend on the position. We can remove this position-dependence from the integration boundary by introducing a new imaginary time coordinate with the transformation $\tau \rightarrow  \tau / (1 + \delta T(\bx)/ T)$. Assuming small temperature variations, we find
\be
S_{\text{coup.}} \simeq \int\text{d}\bx \int_0^{\hbar\beta}\!\!\text{d}\tau \bigg(1 - \frac{\delta T(\bx)}{T}\bigg) A_i(\bx,\tau)J_e^i(\bx,\tau), \label{eq:magex2}
\ee
where we expanded in $\delta T(\bx)/T$. Writing $\delta T(\bx) = x^j\partial_j T(\bx)$ and using from \eqqref{eq:metric} that $\delta g_{j0} = -e^{-i\w \tau}\partial_jT /i\w T$, we can write the second term in \eqqref{eq:magex2} as
\be
\!\!\!\delta S_{\text{coup.}} \simeq \int\text{d}\bx\int_0^{\hbar\beta}\!\!\text{d}\tau A_i(\bx,\tau) J_e^i(\bx,\tau)x^j\delta \dot{g}_{j0}(\bx,
\tau), \!\!\label{eq:magex3}
\ee
where the dot on $\delta g_{j0}$ denotes an imaginary time derivative. If we now use \eqqref{eq:magex3} to calculate the effective action of the gauge field in linear-response theory, we obtain a quadratic contribution already at first order. This diamagnetic-like term exactly constitutes a contribution due to the orbital magnetization, as can be seen from the definition for the magnetization density due to a current density $\bJ_e(\bx)$, i.e.,
\ba
{\bm M}^{\text{orb}}_e = \frac{1}{2V}\int \text{d}\bx \, \langle \bx \times \bJ_e(\bx) \rangle_0, \label{eq:magclas}
\ea
which implies that $\langle J^i_e(\bx,\tau) x^j\rangle_0 \propto \varepsilon^{ijk} M_{e,k}^{\text{orb}}$. A similar line of reasoning can be followed when starting from an action like \eqqref{eq:magex1} with the coupling $J^i_{\veps}\delta g_{i0}$. We find
\be
\delta S_{\text{coup.}} \simeq \int\text{d}\bx\int_0^{\hbar\beta}\!\!\text{d}\tau \,\delta g_{i0}(\bx,\tau) J_{\veps}^i(\bx,\tau)x^j\delta \dot{g}_{j0}(\bx,\tau),
\ee
which can be recognized as a contribution coming from the so-called energy magnetization density, that is defined by replacing the electric current density in \eqqref{eq:magclas} by the energy current density, i.e.,
\ba
{\bm M}_{\veps}^{\text{orb}} = \frac{1}{2V}\int \text{d}\bx \, \langle \bx \times \bJ_{\veps}(\bx) \rangle_0. \label{eq:magclas}
\ea
As we only considered the contributions to the currents due to the current-current correlators in Sec.\ref{sec:currentcurrent}, we need to add the diamagnetic-like contributions coming from the (heat) orbital magnetization. These contributions are only non-zero in the case of broken time-reversal symmetry. In our case this is provided for by the tilting direction $\bt$. As we shall see, the magnetization densities point in the opposite direction of the tilt. 
%%%%%%%%%%%%%%%%%%%%%%%%%%%%%%%%%%%%%%%%%%%%%%%%%%%%%%%%%%%%%%%%%%%%%%%%%%%%%%%%%%%%%%%%%%%%%%%%%%%%%%%%%%%%%%%%%%%%%%%%%%%%%%%%%%%%%%%%%%%%%%%%%%%%%%%%%%%%%%%%%%%%%%%%%%%%%%%%%%%%%%%%%%%%%%%%%%%%%%%%%%%%%%%%%%%%%%%%%%%%%%%%%%%%%%%%%%%%%%%
\subsection{Orbital magnetization due to tilted cones}
A convenient way to calculate the (heat) orbital magnetization is by expressing them in terms of Bloch wavefunctions\cite{Thonhauser2011}. This follows from the semiclassical theory of Bloch electron dynamics, in which electrons can be described as wave packets that are constructed by forming a superposition of the Bloch states of a band \cite{Niu2010}. Such a wave packet has a non-zero spread in real space, such that it can rotate around its center of mass, leading to an orbital magnetic moment. For a band with Bloch wavefunction $|u_{n\chi\bk}\rangle$, the orbital magnetic moment is given by $e{\bm m}^{(1)}_n(\bk)$, with\cite{Niu1999}
\be
{\bm m}^{(p)}_n(\bk) = -\frac{i}{2}\langle \partial_{\bk} u_{n\chi\bk}|\times (\mathcal{H}(\bk) - \veps_{n\bk})^p|\partial_{\bk} u_{n\chi\bk} \rangle. \label{eq:mi}
\ee
An explicit calculation for the two-band model from \eqqref{eq:Ham} yields for a cone with chirality $\chi$: $e{\bm m}^{(1)}_{n\chi}(\bk) =-ev_F\chi \bk/2|\bk|^2$.
Similar to the way in which microscopic spins add up to form a macroscopic magnetization of a material, the orbital magnetic moment contributes to the macroscopic orbital magnetization ${\bm M}_e^{\text{orb}}$. However, besides the contribution of the orbital magnetic moment, there is also a contribution from the center-of-mass motion of the wave packet. The total temperature-dependent orbital magnetization density can therefore be expressed as\cite{Niu2010}
\ba
{\bm M}^{\text{orb}}_e &= \sum_{n,\chi = \pm} \int_{\bk} \bigg[ e{\bm m}^{(1)}_{n\chi}(\bk)\NFD{\veps_{n\bk} - \mux} \nonumber\\
&\phantom{=}+ ek_BT {\bm \Omega}_{n\chi}(\bk)\log\Big(1 + e^{-\beta(\veps_{n\bk} - \mux)}\Big)\bigg], \label{eq:Morb}
\ea
In this expression the first term is simply a thermodynamic average over the orbital magnetic moments and thus ultimately due to the self-rotation of the wave packet, whereas the second term is due to the center-of-mass motion of the wave packet.

Recall that the Weyl nodes in our model are located at the same position in momentum space, because we are interested in intrinsic contributions to the various conductivities, as opposed to topological contributions. Therefore, the only vector that contributes to ${\bm M}_e^{\text{orb}}$ is the tilting direction $\bt$, such that we can write ${\bm M}^{\text{orb}}_e = M^{\text{orb}}_e\bt$. We perform the integral in \eqqref{eq:Morb} by going to spherical coordinates and subtract the Dirac sea, yielding
\ba
M^{\text{orb}}_e &= -\frac{e}{8\pi^2v_F}\sum_{\chi=\pm}\frac{\chi t_{\chi}}{(1-t^2)t} \bigg[ \mux^2 + \frac{\pi^2k_B^2T^2}{3} \bigg] \nonumber\\
&\equiv  \sum_{\chi = \pm} M^{\text{orb}}_{e,\chi}.
\ea
In a completely similar fashion we calculate the circulating contribution in the heat current. Analogously to the electric orbital magnetization, the heat orbital magnetization can be expressed as\cite{Niu2011}
\ba
&{\bm M}_{Q}^{\text{orb}} = \sum_{n,\chi=\pm}\int_{\bk} {\bm \Omega}_{n\chi}(\bk)\int_0^{\veps_{n\bk} - \mux}\!\text{d}x\,x \NFD{x} \\
&\phantom{=}\!\!\!\!-\int_{\bk} \big[(\veps_{n\bk}-\mux){\bm m}_{n\chi}^{(1)}(\bk)+{\bm m}_{n\chi}^{(2)}(\bk)/4\big]\NFD{\veps_{n\bk} -\mux}\nonumber.
\ea
Using \eqqref{eq:mi} we find ${\bm m}^{(2)}_{n\chi}(\bk) = n\chi v_F^2\bk/|\bk|$. Again writing ${\bm M}^{\text{orb}}_{Q} = M^{\text{orb}}_{Q}\bt$ we then find for the finite contribution to the heat orbital magnetization density
\ba
M^{\text{orb}}_{Q} &=-\sum_{\chi=\pm}\frac{\chi\mux\big[(2-t^2)\mux^2  + t^2\pi^2 k_B^2T^2\big]}{24\pi^2(1-t^2)^2v_F} \nonumber \\
&\equiv \sum_{\chi = \pm} M^{\text{orb}}_{Q,\chi} .
\ea
Now that we have explicitly calculated the diamagnetic-like orbital magnetization contributions due to the tilting of the cones, we calculate the contributions to the current coming from the current-current response functions.
%%%%%%%%%%%%%%%%%%%%%%%%%%%%%%%%%%%%%%%%%%%%%%%%%%%%%%%%%%%%%%%%%%%%%%%%%%%%%%%%%%%%%%%%%%%%%%%%%%%%%%%%%%%%%%%%%%%%%%%%%%%%%%%%%%%%%%%%%%%%%%%%%%%%%%%%%%%%%%%%%%%%%%%%%%%%%%%%%%%%%%%%%%%%%%%%%%%%%%%%%%%%%%%%%%%%%%%%%%%%%%%%%%%%%%%%%%%%%%%%%%%%%%%%%%%%%%%%%%%%%%%%%%%%%%%%%%%%%%%%%%%%%%%%%%%%%%%%%%%%%%%%%%%%%%%%%%%%%%%%%%%%%%%%%%%%%%%%%%%%%%%%%%%%%%%%%%%%%%%%%%%%%%%%%%%%%%%%%%%%%%%%%%%%%%%%%%%%%%%%%%%%%%%%%%
\subsection{Electric, energy and mixed current-current correlators}
We start by considering the current-current response functions $\Pi^{k}_{ee}(\w,\bq;\bt)$, $\Pi^{k}_{e\veps}(\w,\bq;\bt)$ and $\Pi^{k}_{\veps\veps}(\w,\bq;\bt)$ that describe the linear response of the electric and energy current densities. Once we have the corresponding transport coefficients, the response for the electric and heat current density can then be obtained by using the relation $\langle \bJ_Q \rangle = \sum_{\chi}\big[\langle \bJ^{\chi}_{\veps} \rangle + (\mux/e)\langle \bJ_e^{\chi}\rangle\big]$. 

In the local, i.e. $\bq = {\bf 0}$, limit, we find using \eqqref{eq:Pigeneric2} together with \eqqref{eq:fee}, \eqqref{eq:fepse}, and \eqqref{eq:fepseps}, 
\be
\frac{i\Pi^k_{ab}(\w^+\!,{\bf 0};\bt)}{2\w v_F^2} =\sum_{\chi,u=\pm}\chi\!\int_{\bk}\frac{h^k_{ab}(\bk,\bt)\NFD{u\veps_{u\bk} + u\mux}}{(\w^+)^2 - 4v_F^2|\bk|^2},
\ee
with $h_{ee}^k(\bk,\bt) =e^2\hat{k}^k$, $h_{e\veps}^k(\bk,\bt) = -ev_F(\bk\cdot\bt)\hat{k}^k$ and $h_{\veps\veps}^k(\bk,\bt) = v_F^2(\bk\cdot\bt)^2\hat{k}^k$. The integrals above have to be proportional to $\bt$, as there is no other vector left. The anomalous Hall conductivity thus follows from $\sahe(\w) = i\Pi^k_{ee}(\w^+\!,{\bf 0};\bt)t_k/\w t^2$. Similarly we define $\alpha^{\text{ANE}}_{\veps}(\w) \equiv i\Pi^k_{e\veps}(\w^+\!,{\bf 0};\bt)t_k/\w t^2$ and $\bar{\kappa}^{\text{THE}}_{\veps}(\w) = i\Pi^k_{\veps\veps}(\w^+\!,{\bf 0};\bt)t_k/\w t^2$, where the subscript `$\veps$' refers to the fact that these linear-response coefficients are for the coupled electric and energy current response and do not contain the magnetization subtractions yet.
 
To obtain the coefficients $\sahe$, $\alpha^{\text{ANE}}_{\veps}$ and $\bar{\kappa}^{\text{THE}}_{\veps}$, we simplify the remaining integral by again choosing spherical coordinates along the direction of $\bt$ like we did to obtain \eqqref{eq:CMCw}. The angular integrals are then easily performed, whereas the integral over $|\bk|$ can only be performed analytically in two specific cases: 1) at $T=0$ for all $\w$ and 2) at $\w = 0$ for all $T$. Here, we focus on the second case because we have only obtained the magnetizations at zero frequency in the previous section. Defining the integrals
\be
I_n(\mux)\equiv \int_0^{\infty} \!\!\text{d}\veps \,\veps^{n-1} \big[\NFD{\veps + \mux} + (-1)^n\NFD{\veps-\mux}\big]
\ee
and
\be
J_n(t)\equiv \int_{-1}^1\!\text{d}y\frac{y^n}{(1+t y)^n},
\ee
we find for the anomalous Hall conductivity
\ba
\s^{\text{AHE}} &= \sum_{\chi=\pm}\frac{\chi e^2t_{\chi}}{8\pi^2t^2v_F}J_1(t)I_1(\mux) \nonumber \\
&=\frac{e^2l(t)}{4\pi^2v_F}\sum_{\chi=\pm}\frac{\chi\mux t_{\chi}}{t} \equiv \sum_{\chi = \pm} \s_{\chi}^{\text{AHE}}. \label{eq:sahe}
\ea
Furthermore we find for $\alpha^{\text{ANE}}_{\veps}$
\ba
\hspace{-.3cm}\alpha_{\veps}^{\text{ANE}} &= \sum_{\chi=\pm}\frac{e\chi t_{\chi}}{8\pi^2Ttv_F}J_2(t) I_2(\mux) \nonumber\\
&=\sum_{\chi=\pm}\frac{e\chi t_{\chi}}{8\pi^2v_FTt}\bigg[\frac{1}{1-t^2} - 2l(t)\bigg]\bigg[\mux^2+ \frac{\pi^2k_B^2T^2}{3}\bigg] \nonumber\\
&\equiv \sum_{\chi = \pm} \alpha_{\veps,\chi}^{\text{ANE}},
\ea
and finally for $\bar{\kappa}^{\text{THE}}_{\veps}$
\ba
&\bar{\kappa}^{\text{THE}}_{\veps} = \sum_{\chi=\pm}\frac{\chi t_{\chi}}{8\pi^2Tv_F}J_3(t)I_3(\mux) \nonumber \\
&=\sum_{\chi=\pm}\frac{\chi t_{\chi}\mux\big(\mux^2 + \pi^2 k_B^2T^2\big)}{4\pi^2v_FTt}\bigg[l(t) - \frac{1-2t^2}{3(1-t^2)^2}\bigg] \nonumber\\
&\equiv \sum_{\chi = \pm} \bar{\kappa}^{\text{THE}}_{\veps,\chi}.
\ea
Upon performing the sum over the two cones with opposite chirality, we find for the anomalous Hall effect $e^2\mu_5l(t)/2\pi^2$ in the case of inversion-symmetry breaking tilt ($t_{\chi} = t$) and $e^2\mu l(t)/2\pi^2$ when inversion symmetry is retained ($t_{\chi} = \chi t$). Furthermore, we note that the anomalous Nernst conductivity is only dependent on temperature in the case of inversion-symmetric tilt. The term proportional to $\mux^2$ in the expression for the anomalous Nernst effect is ill-defined in the low-temperature limit and should be compensated for when we subtract the orbital magnetization density. 
%%%%%%%%%%%%%%%%%%%%%%%%%%%%%%%%%%%%%%%%%%%%%%%%%%%%%%%%%%%%%%%%%%%%%%%%%%%%%%%%%%%%%%%%%%%%%%%%%%%%%%%%%%%%%%%%%%%%%%%%%%%%%%%%%%%%%%%%
%%%%%%%%%%%%%%%%%%%%%%%%%%%%%%%%%%%%%%%%%%%%%%%%%%%%%%%%%%%%%%%%%%%%%%%%%%%%%%%%%%%%%%%%%%%%%%%%%%%%%%%%%%%%%%%%%%%%%%%%%%%%%%%%%%%%%%%%%%%%%%%%%%%%%%%%%%%%%%%%%%%%%%%%%%%%%%%%%%%%%%%%%%%%%%%%%%%%%%%%%%%%%%%%%%%%%%%%%%%%%%%%%%%%%%%%%%%%%%%%%%%%%%%%%%%%%%%%%%%%%%%%%%%%%%%%
\subsection{Thermoelectric transport coefficients by subtraction}
Now that we have obtained explicit expressions for the orbital magnetizations, we can compute the transport coefficients. For the anomalous Nernst effect we find,
\ba
\alpha^{\text{ANE}}T &= \sum_{\chi=\pm} \bigg[\alpha^{\text{ANE}}_{\veps,\chi}T +\frac{\mux}{e}\s^{\text{AHE}}_{\chi} + M^{\text{orb}}_{e,\chi}\bigg] \nonumber\\
&= -\frac{ek_B^2T^2l(t)}{12v_F\hbar^2}\sum_{\chi=\pm}\frac{\chi t_{\chi}}{t}, \label{eq:ANE}
\ea
where we reinstated $\hbar$. For small $t$ this result coincides with results obtained elsewhere\cite{Bardarson2017B}. Due to the subtraction of the orbital magnetization the anomalous Nernst coefficient is now well-behaved in the limit $T\rightarrow0$. Additionally, we note that in the case of a tilt that breaks inversion symmetry, i.e., $t_{\chi} = t$, the contributions from the two cones with opposite chirality subtract, yielding zero. In the case of inversion-symmetry preserving tilt and no chiral imbalance, it is easy to see from \eqqref{eq:sahe} and \eqqref{eq:ANE} that the Mott-like rule
\be
\alpha^{\text{ANE}} = -\frac{\pi^2k_B^2T}{3e}\frac{\text{d}\sahe(\mu)}{\text{d}\mu}, \label{eq:Mott}
\ee
which was derived by Niu \emph{et al.}, is satisfied\cite{Niu2006}. Note that there is a relative sign in \eqqref{eq:Mott} because we have defined the anomalous Nernst current as $\langle \bJ_e \rangle = \alpha^{\text{ANE}}\bt\times {\bm \nabla T}$, i.e., with the same overall sign as the anomalous Hall current $\langle \bJ_e \rangle = \sahe \bt\times\bE$.

Similarly, we compute the coefficient $\bar{\kappa}^{\text{THE}}$ by combining the results from the current-current correlators and the heat orbital magnetization, finding
\ba
\bar{\kappa}^{\text{THE}} T &= \sum_{\chi=\pm} \bigg[\kappa_{\chi}T +2\frac{\mux}{e}\alpha_{\chi}T + \frac{\mux^2}{e^2}\s^{\text{AHE}}_{\chi} + 2M^{\text{orb}}_{Q,\chi}\bigg] \nonumber\\
&=\frac{l(t)k_B^2T^2}{12v_F\hbar^2}\sum_{\chi=\pm}\frac{\chi \mux t_{\chi}}{t}, \label{eq:THE}
\ea
where we also reinstated $\hbar$. From the expression above it is clear that $\bar{\kappa}^{\text{THE}}$ is only non-zero within our simple model when either inversion symmetry is broken and there is a chiral imbalance $\mu_5$, or when inversion symmetry is retained and there is a non-zero chemical potential $\mu$. In the latter case we find for the thermal Hall coefficient from \eqqref{eq:kappaintro},
\ba
\kappa^{\text{THE}} &= \frac{l(t)\mu k_B^2T}{6v_F\hbar^2}\bigg[1 - \frac{1}{3}\frac{k_B^2T^2}{\mu^2}\bigg].
\ea 
In the low-temperature limit $ k_BT/\mu \ll 1$ the second term is negligible and $\kappa^{\text{THE}}$ reduces to previously obtained results\cite{Bardarson2017B}. Additionally, we observe that in this limit the Wiedemann-Franz law
\be
\kappa^{\text{THE}} = \frac{\pi^2k_B^2}{3e^2}T \s^{\text{AHE}},
\ee
for the off-diagonal transport coefficients, is obeyed.
%%%%%%%%%%%%%%%%%%%%%%%%%%%%%%%%%%%%%%%%%%%%%%%%%%%%%%%%%%%%%%%%%%%%%%%%%%%%%%%%%%%%%%%%%%%%%%%%%%%%%%%%%%%%%%%%%%%%%%%%%%%%%%%%%%%%%%%%%%%%%%%%%%%%%%%%%%%%%%%%%%%%%%%%%%%%%%%%%%%%%%%%%%%%%%%%%%%%%%%%%%%%%%%%%%%%%%%%%%%%%%%%%%%%%%%%%%%%%%%%%%%%%%%%%%%%%%%%%%%%%%%%%%%%
\section{Conclusion and discussion} \label{sec:discussion}
In this paper we have investigated the off-diagonal linear response of a pair of tilted Weyl cones, when subjected to a temperature gradient, electric field, magnetic field and vorticity. We focussed on the electronic contributions to the electric and heat current densities and neglected contributions from e.g. \!phonons. As the off-diagonal response is inherently dissipationless, we considered a clean system without disorder. In addition, we neglected the influence of Coulomb interactions among the Weyl fermions. Finally, to preserve clarity, we concentrated on the tilt dependence and did not take the well-known topological contribution to the anomalous Hall and thermal Hall effect into account by taking the the momentum-space separation between the Weyl nodes equal to zero. Under these assumptions, we performed linear-response theory and calculated the appropriate current-current response functions. As the off-diagonal response is determined by their antisymmetric part, we explicitly showed how this part of the current-current response functions can be decomposed in terms of the tilting direction and the external wavenumber.

In the case of the chiral magnetic conductivities, we found that the response remains isotropic when considering tilted cones. In the static limit these conductivities even remain universal, which can be attributed to the chiral anomaly. In the homogeneous or transport limit, on the other hand, the conductivities are renormalized. The situation turned out to be very different for the vortical conductivities: these are generically anisotropic and can be decomposed into a component longitudinal and transverse to the tilting direction. The corresponding longitudinal and transverse chiral vortical conductivities have different values in the static and homogeneous limit as well, but are always tilt-dependent and thus never universal. 

To verify our results coming from the Kubo formalism, we used a combination of exact quantum and semiclassical arguments, thereby explaining the universality of the chiral magnetic conductivities. In the case of the anisotropic vortical conductivities we argued that already simple integrals over the anomalous velocity due to a non-zero Berry curvature, necessarily become anisotropic when including a tilting direction. We were, however, not able to simply explain the appropriate expressions for the effective magnetic field due to vorticity in the case of non-zero tilt and plan to investigate this in future work.

Moreover, we calculated the magnetovortical transport coefficients and the susceptibility not only in the static and homogeneous limit, but also in the more general long-wavelength limit. This turned out to be another source of anisotropy, as the results depend on the angle between the tilting direction and the external wavenumber. To illustrate the effect of this anisotropy we computed the dispersion relation of the chiral magnetic wave using the long-wavelength result for the susceptibility. Interestingly, we found that there is a significant dependence of the chiral wave velocity on the angle between the external magnetic field and the tilting direction. Next to this angle-dependence, we showed that the wave becomes soft when the tilt of the cone becomes too large, signaling the Lifshitz transition from a type-I to a type-II Weyl cone. For zero tilt, however, we found that using simply the static-limit result for the susceptibility is a rather good approximation.

In addition, we showed that also the frequency dependence of the chiral magnetic and the chiral vortical conductivities is very different indeed. The AC chiral magnetic conductivity is unusual in the sense that at zero tilt, its imaginary part is given by delta functions centered around $\w = \pm2\mu_{\pm}$. Ultimately this is due to the fact that in order to obtain the part of the electric current-current response function that is linear in the external wavenumber, we needed to expand a propagator, thereby turning the bubble diagram into a triangle diagram. For a non-zero tilt, however, the imaginary part attains a finite height and width. The behavior of the AC vortical conductivities was rather different, as the appurtenant current-current response functions were already linear in the external wavenumber. The imaginary part is therefore determined by the more usual Heaviside step functions and is finite both for zero and non-zero tilt. 

In the last part of this paper we concentrated on the off-diagonal thermoelectric transport and elucidated how magnetization contributions to the current occur as diamagnetic-like contributions. Subsequently, we obtained the contribution from the magnetization explicitly by performing a microscopic calculation. As it turns out, the magnetizations are always pointing in the opposite direction of the tilting direction. Having obtained the magnetizations explicitly, we subtracted them to find the transport linear-response coefficients. We found it an illustrative exercise to do this explicitly and it would be interesting to investigate how this scheme can be extended to non-zero frequencies. An important difference with the magnetovortical coefficients turned out to be that the thermoelectric coefficients are odd functions of the tilting direction, whereas the former are even functions of the tilting direction. The magnetovortical coefficients therefore do not depend on whether the tilt breaks inversion symmetry or not. Contrastingly, in the case of the thermoelectric coefficients, the anomalous Hall and thermal Hall effect are only non-zero in the case of broken inversion symmetry if there is a chiral imbalance. The anomalous Nernst effect is only non-zero when inversion symmetry is retained. 

For future research it would interestering to investigate the influence of disorder and Coulomb interactions on the magnetovortical conductivities, which has already been explored for the chiral magnetic conductivity\cite{Sau2016}.

\section*{ACKNOWLEDGMENTS}
This work is supported by the Stichting voor Fundamenteel Onderzoek der Materie (FOM) and is part of the D-ITP consortium, a program of the Netherlands Organisation for Scientific Research (NWO) that is funded by the Dutch Ministry of Education, Culture and Science (OCW).
\medskip
\bibliographystyle{apsrev4-1}
\bibliography{Bibliography}
%%%%%%%%%%%%%%%%%%%%%%%%%%%%%%%%%%%%%%%%%%%%%%%%%%%%%%%%%%%%%%%%%%%%%%%%%%%%%%%%%%%%%%%%%%%%%%%%%%%%%%%%%%%%%%%%%%%%%%%%%%%%%%%%%%%%%%%%%%%%%%%%%%%%%%%%%%%%%%%%%%%%%%%%%%%%%%%%%%%%%%%%%%%%%%%%%%%%%%%%%%%%%%%%%%%%%%%%%%%%%%%%%%%%%%%%%%%%%%%%%%%%%%%%%%%%%%%%%%%%%%%%%%%%%%%%%%%%%%%%%%%%%%%%%%%%%%%%%%%%%%%%%%%%%%%%%%%%%%%%%%%%%%%%%%%%%%%%%%%%%%%%%%%%%%%%%%%%%%%%%%%%%%%%%%%%%%%%%%%%%%%%%%%%%%%%%%%%%%%%%%%%%%%%%%%%%%%%%%%%%%%%%%%%%%%%%%%%%%%%
\vspace{-.1cm}
\appendix
\onecolumngrid
\section{Explicit expressions} \label{app:explicitexpressions}
\noindent In this Appendix we give some explicit expressions that were too lengthy to put in the main text. We start with the functions $f^{k,uv}_{ab}(\bq,\bt,\bk)$ that we used in \eqqref{eq:Pigeneric2}. For $f^{k,uv}_{\veps p}(\bq,\bt,\bk)$ and $f^{k,uv}_{ee}(\bq,\bt,\bk)$ the full dependence on $\bq$ is tractable. They are given by 
\ba
f^{k,uv}_{\veps p}(\bq,\bt,\bk) &=  -\frac{\big[u|\bk| + v|\bk+\bq| + 2uv(\bk\cdot\bt) + uv(\bq\cdot\bt) \big](2\bk + \bq)\cdot\big(\bq\,k^k - \bk\,q^k\big)}{4|\bk||\bk+\bq|}, \label{eq:fepsp}\\
f^{k,uv}_{ee}(\bq,\bt,\bk) &= 2e^2\bigg[\bigg(\frac{u}{|\bk+\bq|} - \frac{v}{|\bk|}\bigg)k^k + u\frac{q^k}{|\bk + \bq|}  + uv \frac{(\bk\cdot\bt)q^k - (\bq\cdot\bt) k^k}{|\bk||\bk+\bq|}\bigg].\label{eq:fee}
\ea
The leading-order expressions for $f^{k,uv}_{\veps e}(\bq,\bt,\bk)$ and $f^{k,uv}_{\veps \veps}(\bq,\bt,\bk)$ are given by
\ba
f^{k,uv}_{\veps e}(\bq,\bt,\bk) &= -e\bigg[2\bigg(\frac{u}{|\bk+\bq|} - \frac{v}{|\bk|}\bigg)(\bk\cdot\bt)k^k  + (3u+v)(\hat{\bk}\cdot\bt)q^k - 2v(\bq\cdot\bt)\hat{k}^k  +(1+uv)q^k - 2uv(\hat{\bk}\cdot\bq)\hat{k}^k \label{eq:fepse}\nonumber\\
&\phantom{=}- 2uv(\hat{\bk}\cdot\bt)(\bq\cdot\bt)\hat{k}^k + 2uv(\hat{\bk}\cdot\bt)^2q^k\bigg] + \mathcal{O}(|\bq|^2), \\
f^{k,uv}_{\veps \veps}(\bq,\bt,\bk) &= -2\bigg(\frac{u}{|\bk+\bq|} - \frac{v}{|\bk|}\bigg)(\bk\cdot\bt)^2k^k + (1+uv)\big[(\bk\cdot\bt)q^k - (\bq\cdot\bt)k^k\big] - 2uv(\hat{\bk}\cdot\bt)^2(\bq\cdot\bt)k^k\nonumber\\
&\phantom{=}+ (u+v)\big(|\bk|q^k+ (\bk\cdot\bq)\hat{k}^k\big) + 2(2u+v)(\hat{\bk}\cdot\bt)^2|\bk|q^k   - 4v(\bk\cdot\bt)(\bq\cdot\bt)\hat{k}^k  - 4uv(\hat{\bk}\cdot\bq)(\hat{\bk}\cdot\bt)k^k  \nonumber\\
&\phantom{=} + 2uv|\bk|(\hat{\bk}\cdot\bt)^3q^k + \mathcal{O}(|\bq|^2) \label{eq:fepseps}.
\ea
From these expressions it becomes clear which terms contribute to the thermoelectric transport coefficients. Indeed, the first two terms in $f^{k,uv}_{ee}(\bq,\bt,\bk)$, $f^{k,uv}_{\veps e}(\bq,\bt,\bk)$ and $f^{k,uv}_{\veps\veps}(\bq,\bt,\bk)$ are the only terms that are non-zero when $|\bq| = 0$. An expansion of $N^{\chi}_{uv}(\w,\bq,\bt,\bk)$ with $uv = -1$ for small $|\bq|$ gives a term proportional to $\w$, as can be seen in \eqqref{eq:Nplusmin}, thereby leading to the terms proportional to $\w t^k$ in Eqs.\,\eqref{eq:DecompPiee}-\eqref{eq:DecompPiepseps}.
\newline \newline
Furthermore, the function $W(x,\bt)$ that we used in the expressions for the chiral magnetic conductivities \eqqref{eq:scmepslwl} and \eqqref{eq:scmelwl}, is defined by
\be
W(x,\bt) \equiv \frac{1-(x-\tpara)^2}{Z(x,\bt)^2}\bigg[1 + \frac{x}{2}Y(x,\bt)\bigg] = \begin{dcases} (1-t^2)^{-1} \quad \text{for} \quad x\rightarrow0, \\ l(t) \qquad \qquad\text{for} \quad x\rightarrow\infty, \end{dcases}
\ee
in terms of 
\begin{subequations}
\ba
Y(x,\bt) &\equiv \frac{1}{Z(x,\tpara,t)} \log\Bigg(\frac{[x-\tpara -1][1 - \tpara + \tpara x - t^2(1+x-\tpara) +  (1-\tpara)Z(x,\bt)]}{[x-\tpara +1][1 + \tpara + \tpara x - t^2(1-x-\tpara) +  (1+\tpara)Z(x,\bt)]}\Bigg), \\
Z(x,\bt) &\equiv \sqrt{1 + 2\tpara x - \tpara^2 -t^2 + t^2(x-\tpara)^2}.
\ea
\end{subequations}
Finally, the explicit expression for the susceptibility in the long-wavelength limit from \eqqref{eq:suscep} can also be expressed in terms of $W(x,\bt)$ in the following way
\be
\chi_{\pm}(x, {\bf 0};\bt) = \frac{\mu_{\pm}^2}{4\pi^2v_F^3Z(x,\bt)^2} \Bigg[ \frac{\tpara^2 - t^2}{1-t^2} +\bigg(2 + t^2 -3\tpara^2 + \frac{2x^2}{1-(x-\tpara)^2}\bigg)W(x,\bt)\Bigg].
\ee
Two useful limiting cases for $\chi_{\pm}(x,{\bf 0};\bt)$ are
\be
\lim_{x\to 0} \chi_{\pm}(x, {\bf 0}, \bt) = \frac{\mu_{\pm}^2}{2\pi^2(1-t^2)^2v_F^3},
\ee
which we used in \eqqref{eq:suscep2}, and
\be
\lim_{\bt \to {\bf 0}} \chi_{\pm}(x, {\bf 0};\bt) = \frac{\mu_{\pm}^2}{2\pi^2v_F^3}\bigg[1 - \frac{x}{2}\log\bigg(\frac{x+1}{x-1}\bigg)\bigg].
\ee
%%%%%%%%%%%%%%%%%%%%%%%%%%%%%%%%%%%%%%%%%%%%%%%%%%%%%%%%%%%%%%%%%%%%%%%%%%%%%%%%%%%%%%%%%%%%%%%%%%%%%%%%%%%%%%%%%%%%%%%%%%%%%%%%%%%%%%%%%%%%%%%%%%%%%%%%%%%%%%%%%%%%%%%%%%%%%%%%%%%%%%%%%%%%%%%%%%%%%%%%%%%%%%%%%%%%%%%%%%%%%%%%%%%%%%%%%%%%%%%%%%%%%%%%%%%%%%%%%%%%%%%%%%%%%%%%%%%%%%%%%%%%%%%%%%%%%%%%%%%%%%%%%%%%%%%%%%%%%%%%%%%%%%%%%%%%%%%%%%%%%%%%%%%%%%%%%%%%%%%%%%%%%%%%%%%%%%%%%%%%%%%%%%%%%%%%%%%%%%%%%%%%%%%%%%%%%%%%%%%%%%%%%%%%%%%%%%%%%%%%%%%%%%%%%%%%%%%%%%%%%%%%%%%%%%%%%%%%%%%%%%%%%%%%%%%%%%%%%%%%%%%%%%%%%%%%%%%%%%%%%%%%%%%%%%%%%%%%%%%%%%%%%%%%%%%%%%%%%%%%%%%%%%%%%%%%%%%%%%%%%%%%%%%%%%%%%%%%%%%%%%%%%%%%%%%%
\section{Frequency dependence magnetovortical conductivities} \label{app:freq}
\noindent Below we list the full frequency dependence of the chiral magnetic and chiral vortical conductivities. They can all be expressed in terms of the functions $L_i(\w)$ defined in Eqs.\eqref{eq:L1}, \eqref{eq:L2} and \eqref{eq:L3}. We find
\ba
\sigma^{\text{CVE}}_{\parallel}(\w) &= -\sum_{\chi = \pm}\frac{\chi \mux^2}{48\pi^2t^3}\bigg[ \frac{(2+t^2)t}{1-t^2} - \frac{(1+3t^2)\w^2 + 12\mux^2}{16\mux^2}L_1(\w) + \frac{3(1+t^2)\w^2 + 4\mux^2}{8\w \mux}L_2(\w) + \frac{\w^2t^3}{4\mux^2}L_3(\w)  \bigg], \\
\sigma^{\text{CVE}}_{\perp}(\w) &= \sum_{\chi = \pm}\frac{\chi \mux^2}{96\pi^2t^3}\bigg[ \frac{(2-5t^2)t}{1-t^2} - \frac{(1-9t^2)\w^2 + 12\mux^2}{16\mux^2}L_1(\w) + \frac{3(1-3t^2)\w^2 + 4\mux^2}{8\w\mux}L_2(\w) - \frac{\w^2t^3}{2\mux^2}L_3(\w)  \bigg], \\
\sigma_{\veps,\parallel}^{\text{CVE}}(\w) &= -\sum_{\chi = \pm}\frac{\chi\mux^3}{64\pi^2t^3}\bigg[\frac{(3t^4+14t^2 - 9)t}{3(1-t^2)^2} - \frac{(1+3t^2)t\w^2}{4\mux^2} + \frac{(1+t^2)\w^2 + 4\mux^2}{4\mux^2}L_1(\w) \nonumber\\
&\phantom{=}\qquad\qquad\qquad\qquad -\frac{(1+2t^2-3t^4)\w^4 + 8(3+t^2)\w^2\mux^2 + 16\mux^4}{32\w\mux^3}L_2(\w) \bigg], \\
\sigma_{\veps,\perp}^{\text{CVE}}(\w) &= -\sum_{\chi = \pm}\frac{\chi\mux^3}{128\pi^2t^3}\bigg[\frac{(9-14t^2+13t^4)t}{3(1-t^2)^2} + \frac{(1-5t^2)t\w^2}{4\mux^2} - \frac{(1-3t^2)\w^2 + 4\mux^2}{4\mux^2}L_1(\w) \nonumber\\
&\phantom{=}\qquad\qquad\qquad\qquad +\frac{(1-6t^2+5t^4)\w^4 + 24(1-t^2)\w^2\mux^2 + 16\mux^4}{32\w\mux^3}L_2(\w) \bigg], \\
\sigma_{\veps}^{\text{CME}}(\w) &= -\sum_{\chi = \pm} \frac{\chi\mux^2(1-t^2)}{24\pi^2t^3}\bigg[\frac{(t^2 - 4)t}{1-t^2} + \frac{\w^2 + 12\mux^2}{8\mux^2}L_1(\w) - \frac{3\w^2 + 4\mux^2}{4\w\mux}L_2(\w) - \frac{\w^2t^3}{8\mux^2}L_3(\w)\bigg].
\ea
%%%%%%%%%%%%%%%%%%%%%%%%%%%%%%%%%%%%%%%%%%%%%%%%%%%%%%%%%%%%%%%%%%%%%%%%%%%%%%%%%%%%%%%%%%%%%%%%%%%%%%%%%%%%%%%%%%%%%%%%%%%%%%%%%%%%%%%%%%%%%%%%%%%%%%%%%%%%%%%%%%%%%%%%%%%%%%%%%%%%%%%%%%%%%%%%%%%%%%%%%%%%%%%%%%%%%%%%%%%%%%%%%%%%%%%%%%%%%%%%%%%%%%%%%%%%%%%%%%%%%%%%%%%%%%%%%%%%%%%%%%%%%%%%%%%%%%%%%%%%%%%%%%%%%%%%%%%%%%%%%%%%%%%%%%%%%%%%%%%%%%%%%%%%%%%%%%%%%%%%%%%%%%%%%%%%%%%%%%%%%%%%%%%%%%%%%%%%%%%%%%%%%%%%%%%%%%%%%%%%%%%%%%%%%%%%%%%%%%%%%%%%%%%%%%%%%%%%%%%%%%%%%%%%%%%%%%%%%%%%%%%%%%%%%%%%%%%%%%%%%%%%%%%%%%%%%%%%%%%%%%%%%%%%%%%%%%%%%%%%%%%%%%%%%%%%%%%%%%%%%%%%%%%%%%%%%%%%%%%%%%%%%%%%%%%%%%%%%%%%%%%%%%%%%%%%
\end{document}